\documentclass[journal]{IEEEtran}
\usepackage[style=ieee,maxnames=4,minnames=3,maxbibnames=3]{biblatex}

\usepackage{amsfonts}
\usepackage[cmex10]{amsmath}
\usepackage{amsthm}

\newcommand{\mr}{\mathrm}

\theoremstyle{definition}

\theoremstyle{remark}

\theoremstyle{remark}

\usepackage{array}
\usepackage{mdwmath}
\usepackage{mdwtab}
\usepackage{eqparbox}
\usepackage[caption=1]{caption}
\usepackage{stfloats}
\usepackage{url}
\usepackage{amssymb}
\usepackage{float}
\usepackage{indentfirst}
\usepackage{makeidx}
\usepackage{tabularx}
\usepackage{cases}
\usepackage{algorithmic}
\usepackage{mathtools}
\usepackage{color}
\usepackage{bm}
\usepackage{stfloats}
\usepackage{lipsum}
\usepackage{amsfonts}
\usepackage{bm}
\usepackage{dsfont}
\usepackage{lipsum}
\usepackage{graphicx}
\usepackage{url}
\usepackage{bbm}
\newcommand{\RNum}[1]{\uppercase\expandafter{\romannumeral #1\relax}}
\ifCLASSOPTIONcompsoc
\usepackage[caption=false, font=normalsize, labelfont=sf, textfont=sf]{subfig}
\else
\usepackage[caption=false, font=footnotesize]{subfig}
\fi
\usepackage{multirow}
\allowdisplaybreaks
\makeatletter

\usepackage[linesnumbered,ruled,vlined]{algorithm2e}
\usepackage{amsmath}
\makeatother

\usepackage{amsthm}
\theoremstyle{definition} 
\theoremstyle{remark} 
\begin{document}
\title{
Generative Diffusion Receivers: Achieving Pilot-Efficient MIMO-OFDM Communications}

        



\author{
    Yuzhi~Yang, Omar~Alhussein, \IEEEmembership{Member,~IEEE,} Atefeh~Arani, \IEEEmembership{Member,~IEEE,}\\ Zhaoyang~Zhang, \IEEEmembership{Senior~Member,~IEEE,} and M\'erouane~Debbah, \IEEEmembership{Fellow,~IEEE,}

\thanks{Y. Yang, O. Alhussein and M. Debbah are with the KU 6G Research Center, College of Computing and Mathematical Sciences, Khalifa University, Abu Dhabi 127788, UAE (e-mails: \{yuzhi.yang, omar.alhussein, merouane.debbah\}@ku.ac.ae). 
}
\thanks{A. Arani is with the Department of Electrical, Computer and Biomedical Engineering, Toronto Metropolitan University, Canada. (e-mail: atefeh.arani@torontomu.ca). 
}
\thanks{Z. Zhang is with the College of Information Science and Electronic Engineering, Zhejiang University, Hangzhou 310027, China, and also with Zhejiang Provincial Key Laboratory of Multi-modal Communication Networks and Intelligent Signal Processing, Hangzhou 310027, China. (e-mail: ning\_ming@zju.edu.cn) 
}
}

\maketitle
\begin{abstract}
This paper focuses on wireless multiple-input multiple-output (MIMO)-orthogonal frequency division multiplex (OFDM) receivers. Traditional wireless receivers have relied on mathematical modeling and Bayesian inference, achieving remarkable success in most areas but falling short in their ability to characterize channel matrices. Neural networks (NNs) have demonstrated significant potential in this aspect. Nevertheless, integrating traditional inference methods with NNs presents challenges, particularly in tracking the error progression. Given the inevitable presence of noise in wireless systems, generative models that are more resilient to noise are garnering increased attention. In this paper, we propose re-evaluating the MIMO-OFDM receiver using diffusion models, which is a common generative approach. With diffusion models, we can effectively leverage prior knowledge of channel matrices and incorporate traditional signal estimation components. Specifically, we explore the diffusion system and introduce an imagination-screening strategy to guide the denoising process. Furthermore, diffusion models enable adaptation to varying noise levels and pilot schemes using the same NN, significantly reducing training and deployment costs. Simulated results reveal that, for pilot densities ranging from 4–6 pilots per 64‑subcarrier block and signal-to-noise ratios (SNRs) from -4 dB to 0 dB, our proposed receiver reduces channel-reconstruction error by up to two times compared to leading deep-learning models, with the most pronounced improvements observed in low-pilot conditions. Additionally, performance enhancements can be achieved with a larger imagination size, despite increased computational complexity.
\end{abstract}

\begin{IEEEkeywords}
Channel estimation, diffusion model, MIMO, OFDM, wireless receiver
\end{IEEEkeywords}

\section{Introduction}
\subsection{Motivation}
Channel estimation is one of the core problems in the multiple-input multiple-output (MIMO)-orthogonal frequency division multiplex (OFDM) receiver. Especially in massive MIMO settings, channel estimation has become the bottleneck of system performance. Traditional modeling and interpolation methods have provided a possible method to reduce pilot overhead, while facing severe performance deterioration with sparse pilots. Neural networks (NNs) later showed great potential in improving this procedure. However, NNs have disadvantages in processing and error prediction with noisy input unless specifically designed and trained. Thus, incorporating NNs with traditional signal processing modules based on Bayesian inference becomes an important problem, where the NN's capability of dealing with noisy input is required.

In a typical MIMO-OFDM receiver, we first estimate the channel given the pilots and then try to recover the transmitted signal from the received signal by the estimated channel.
This classical method has worked very well and is almost optimal except that the prior distribution of the channel is hard to describe explicitly. Therefore, traditional methods usually allocate too many resources for pilots due to the imprecise description of the prior distribution.

On the other hand,  advancements in generative artificial intelligence, illustrated by the diffusion model \cite{ddpm, ddim}, have led to a fresh re-evaluation of traditional challenges through an innovative generative lens, thus enhancing performance. Generative models facilitate the production of high-quality structural data from noisy initialization or even pure noise, with the process being steerable through specific conditions. Diffusion models have achieved remarkable success in image generation tasks, enabling them to create \cite{ddpm, ddim}, satisfy \cite{xu2023versatile}, or edit \cite{kawar2023imagic} high-quality images guided by text which can be extended to other domains involving structural data, such as wireless networks.
We note that, in traditional receiver designs, we usually utilize an iterative algorithm, gradually improving the estimation quality, which is similar to diffusion models. Such similarity inspired us to reconsider the receiver from the perspective of generative models.

In typical diffusion models, incremental noise is introduced during the forward phase until the input nearly becomes random noise. During the reverse phase, NNs systematically remove this noise with the aid of given conditions. From this perspective, the receiver can be re-envisioned as a procedure where it consistently endeavors to generate a channel, informed by learned structural knowledge, while the received signal and the prior knowledge of the transmitted signal serve as guiding conditions. The recovery task then is to construct the channel-transmitted signal pair such that the channel reflects the channel structure found in the dataset, the transmitted signal aligns with the prior knowledge of the data, and the distance between the predicted receiving signal and its practical value is minimized.

We note that there are still notable differences, such that we cannot directly apply the mature algorithms in text-guided diffusion models for image generation. The first and most important difference is that we do not need any imagination in a wireless communication system. Generative models usually have strong capability to generate various high-quality outputs given the same hint.
{In the context of generative models, we usually call such capability ``imagination'' or ``creativity.'' On the contrary, if the model generates images that do not correspond to the condition, it becomes ``hallucination.''
In wireless communication, since we always have a ground truth, the imagination and creativity almost turn to be harmful hallucination, requiring methods to restrict the exaggerated creativity of the model.} Another difference is that we already have strong methods and criteria to accurately evaluate the generated channels, from which we can greatly benefit.

In this paper, we use traditional signal processing modules to provide guidance for the channel-generating NN.
{Through traditional signal processing techniques like semi-blind estimation, we can also provide some gradient guidance based on symbol prior. However, such prior is always weak compared to the original estimation. As the edge information provided by data symbols is always insufficient compared to the direct information from pilots, semi-blind methods usually require far more data symbols than pilots to obtain performance gain. Thus, the gradient guidance by such methods is usually extremely weak compared to the channel prior and pilot-based estimation, and is likely to mislead the denoising process, which is also proven through the negative results in Appendix \ref{app}. Thus, we use it as a selection criteria instead of direct guidance.}
Specifically, we attempt to generate a bunch of channel matrices gradually in each iteration, which are input into the signal processing modules for signal recovery. Referring to the list-voting mechanism in decoding, we then retain several best results for further generation and discard others. Through sufficient iterations, we can eventually obtain an estimation that meets the error requirements.

\subsection{Contributions}
In this paper, we propose a diffusion-based wireless receiver framework for MIMO-OFDM scenarios. Instead of the traditional channel estimation-signal recovery scheme, we turn to utilizing the signal's prior knowledge to screen the channel generation results and use the recovery error as the criterion. Furthermore, with the diffusion model, we no longer need to train different models for different pilot/modulation schemes and channel conditions.
The main contributions of this paper are summarized as follows.
\begin{itemize}
\item We propose a MIMO-OFDM channel generation framework based on the diffusion model. We also propose an imagination and screening method with recovery error as the evaluation criterion for denoising process based on traditional signal processing methods.
\item The proposed scheme is adaptive to different pilot schemes. Through one-time pretraining, the trained NN can be directly applied to various different pilot and modulation schemes, greatly saving retraining cost when the channel state varies.
\item  We also discuss the effect of different parameters on the proposed scheme both analytically and numerically. Simulation results show that the proposed scheme works well under different channel conditions and communication schemes, which means that it is possible to adaptively adjust the communication scheme within the proposed framework.
\end{itemize}
\subsection{Paper Organization}
The paper is organized as follows. Section I is a general introduction, and we provide a brief survey of related works in Section II. The problem considered is explicitly described in Section III, and the proposed framework and the algorithm workflow are shown in Section IV. Section V provides the numerical results. Finally, Section VI concludes the paper and provides some insights for future work.

\section{Related Works}
In this section, we focus on the related works in three aspects. We first summarize the existing NN-based applications in wireless transceivers, which provide reference to the backbone NN framework design of this paper. Further, we exploit the existing works applying diffusion models to communication networks, which are the most related works of this paper. Lastly, we briefly overview the conditional diffusion models. Although these works are irrelevant to the communication system, they are still important references in the algorithm design.
\subsection{NNs in Wireless Transceiver}
With the development of AI, its application in wireless transceivers has received much attention \cite{8054694, 11017513, zhu2025wirelesslargeaimodel}. Within various problems in wireless transceivers, channel characterization problems have become one of the focuses. Unlike many signal processing problems that have a strict mathematical background and where traditional inference-based methods can work well, MIMO-OFDM channel-related problems usually require accurate channel modeling, which is hard to explicitly describe in practice and thus becomes suitable for NNs. Meanwhile, with the development of massive MIMO and OFDM, channel matrices have become extremely large, making it unaffordable to directly estimate channels by pilots. Given the fact that scatterers are always sparsely distributed in space, the degree of freedom in the MIMO-OFDM channels does not increase linearly as the problem size increases. However, some small objects and uncontrollable disturbances may also affect the channels, especially when millimeter waves are used, making it extremely hard to establish an accurate and explicit model. NNs are good at modeling such implicit but highly structural relationships, which provide feasibility for channel compression, retrieval, and mapping with sparse or inaccurate estimations.

Pioneered by CsiNet \cite{CSInet}, NNs have shown incredible capability in the channel feedback problem, showing that the channel can be severely compressed while maintaining retrieval accuracy. Such methods remarkably outperform traditional interpolation-based methods, thus becoming a potential technique for future wireless networks. Some following works have also improved the structure of the NN \cite{9931713, nolonger, guo2025promptenabledlargeaimodels, zhuang2025extractbestdiscardrest}. In addition, similar ideas have been applied to other channel-related tasks. Instead of simple retrieval, similar NNs can also be used to predict uplink channels by downlink ones in frequency division duplex systems \cite{9175003, 10598833}. Through recurrent NNs \cite{RNN-pred, ODERNN} and neural ordinary differential equations \cite{ODE}, we can predict wireless channels from the UE's location.
Moreover, similar techniques can also be applied to prediction \cite{chen2025analogicallearningcrossscenariogeneralization, jiang2025aicsiprediction5gadvanced}. With large AI models, we can also conduct channel prediction in delay-Doppler domain \cite{xue2025largeaimodeldelaydoppler}. However, due to the unavoidable error in localization and the subtle changes in the environment, it is infeasible to directly generate a channel prediction without pilots to an acceptable quality. A more practical idea is to integrate estimation with prediction. Provided some estimated parts of the channel, recent works demonstrate that we can obtain high-quality channel estimations from rough and incomplete ones \cite{channel_mapping, mixer, 10845822}. These works investigate different types of NNs to capture channel characteristics, which provides an important reference for the design of the NN structure in this paper.

AI-based overall transceiver designs are also well investigated, mainly realized by deep unfolding methods. The first such work to rely on deep unfolding is the unfolded iterative shrinkage thresholding algorithm for compressive sensing \cite{LISTA}. In addition, by gradient-based finetuning, deep unfolding has been applied in signal processing algorithms, such as channel estimation for OFDM \cite{8682639}, MIMO precoding \cite{9246287} and detection \cite{10306259}, and integrated sensing and communication (ISAC) transceiver design \cite{11005673}.
Further systematic works have constructed the joint channel and data estimation problems in the wireless receiver \cite{10978568, 10938858, li2025jointchannelestimationsignal}.
Also in our recent work \cite{10942479}, we investigated how to integrate NNs into the traditional signal processing progress. However, prediction error remains a significant challenge.

Meanwhile, industrial efforts have also been made to use such NN-based wireless transceiver applications. DeepMIMO \cite{deepmimo} dataset provides MIMO-OFDM channel matrices in various scenarios generated by the ray-tracing method. Through similar methods, there are also similar datasets such as WAIR-D \cite{wair} and Sionna \cite{sionna} by Nvidia, which also supports radio maps for some MAC layer applications. Such datasets have become the basis for channel-related research. Recently, Nvidia has also carried out the AI-RAN platform \cite{AIRAN}, providing a systematic solution to AI-based wireless transceiver problems.

\subsection{Diffusion Models in Wireless Networks}
\begin{table*}[]\setlength{\tabcolsep}{1pt}
\begin{tabular}{|c|c|c|c|c|}
\hline
Application                               & Diffusion Variable                & Condition                & Comments                                                         & Reference \\ \hline
Survey                                    & -                                 & -                        &                           -                                     & \cite{10529221, 10628027, 10990238, fan2025generativediffusionmodelswireless} \\ \hline
\multirow{3}{*}{\shortstack{Radio Map\\ Reconstruction}} & \multirow{3}{*}{Radio Map}        & N/A                      & Generating radio maps with scenario illustration                 &     \cite{qiu2023irdm, fu2025ckmdiffgenerativediffusionmodel, wang2025radiodiffk2helmholtzequationinformed}    \\ \cline{3-5}
                                          &                                   & Partial Observations     & Using sparse observations to guide the reconstruction            &    \cite{sortino2024radiff, luo2024rm}  \\ \cline{3-5}
                                          &                                   & Time-Related Information & Importing historical observations for better generation          &     \cite{wang2024radiodiff}  \\ \hline
Signal Recovery                           & Raw Signal                        & N/A                      & Using diffusion models to generate/complete wireless signals     &    \cite{chi2024rf}\\ \hline
ISAC                                      & Scenario Illustration             & Channel Estimations      & Using diffusion models to help sensing in ISAC systems           &    \cite{xu2025brownian, 10839236, 11004012}    \\ \hline
Metaverse                                 & Any                               & Any                      & Diffusion-based Metaverse applications in communication networks &    \cite{10981610, 11018804, wen2025diffusionbaseddynamiccontractfederated}  \\ \hline
\multirow{4}{*}{\shortstack{Wireless\\ Transceiver}}     & \multirow{3}{*}{Wireless Channel} & N/A                      & Generating from a rough channel estimation                       &     \cite{xu2025brownian, ma_diffusion_2024} \\ \cline{3-5}
                                          &                                   & Pilot-Based Estimation   & Using estimation results based on pilots to guide the diffusion  &   \cite{balevi2020high, MIMOdiffusion, fesl_diffusion-based_2024, bhattacharya2025successive} \\
                                          \cline{3-5} &&{Scenario Label}   & {Using learned scenario information to guide the diffusion}  &  \cite{li2025wirelesschannelidentificationconditional, ni2025conditionaldiffusionmodeldrivengenerative}\\ \cline{2-5}& {Semantic Commun.} & {Semantic information} & {Using diffusion models for semantic recovery}
                                          &\cite{grassucci2023generative,wu2024cddm,grassucci2024diffusion,10896580}\\
                                          \hline
\end{tabular}
\caption{Existing diffusion model applications in wireless networks}\label{table::intro}
\end{table*}
Recently, diffusion models have also been used in tasks related to wireless networks \cite{10529221, 10628027, 10990238}. A hot topic is the generation of radio maps from the scenario using diffusion models \cite{qiu2023irdm, fu2025ckmdiffgenerativediffusionmodel}. Based on some samples \cite{sortino2024radiff, luo2024rm} or time-related information \cite{wang2024radiodiff}, conditional diffusion models can be used to generate high-quality radio maps and assist resource allocation algorithms. Besides, with physically informed NN structures, we can improve the generation quality of radio maps \cite{wang2025radiodiffk2helmholtzequationinformed}. Meanwhile, diffusion models can also be used to enhance the generation and supplementation of radio frequency signals \cite{chi2024rf}.
There are also some other works considering the diffusion models in ISAC scenarios \cite{10839236, 11004012}, Metaverses \cite{10981610, 11018804, wen2025diffusionbaseddynamiccontractfederated}. These works regarding the applications of diffusion models in wireless networks are also summarized in Table \ref{table::intro}.

{\begin{table*}[]
\setlength{\tabcolsep}{5pt}
\centering

\begin{tabular}{|c|c|>{\centering\arraybackslash}p{2cm}|>{\centering\arraybackslash}p{9cm}|}
\hline
Working Domain                                 & Function                                                           & Related Works                          & Characteristics                                                                                                                                                        \\ \hline
\multirow{2}{*}{Signal}                        & \multirow{2}{*}{Capturing signal prior}                            &                \cite{grassucci2023generative,wu2024cddm,grassucci2024diffusion,10896580}                        & Use diffusion models to denoise semantic signals; Overlooking detailed channel structure                                                                              \\ \cline{3-4} 
                                               &                                                                     & \multirow{2}{*}{\cite{zilberstein_joint_2024, bhattacharya2025successive}}                      & Simultaneously generate MIMO channel and transmitted symbols through                                       \\ \cline{1-2}
\multirow{3}{*}{Channel}                       & \multirow{2}{*}{Capturing MIMO channel prior}                      &                                        &                                                                                                                                                                    diffusion models; Require complete initial rough estimation    \\ \cline{3-4} 
                                               &                                                                    &        \cite{balevi2020high, MIMOdiffusion,ma_diffusion_2024,fesl_diffusion-based_2024,xu2025brownian,li2025wirelesschannelidentificationconditional, ni2025conditionaldiffusionmodeldrivengenerative}                                 & Requires full initial estimate with relatively poor quality; Signal processing results used for initialization/guidance                                               \\ \cline{2-4} 
                                               & Capturing OFDM channel prior                                       & This paper                             & Simultaneous denoising and completion; Traditional signal processing modules act as diffusion model replicas                                                           \\ \hline
\end{tabular}
\caption{Comparison of Diffusion Model Applications in Wireless Transceivers}\label{table::intro2}
\end{table*}}
As for the wireless transceiver design, {an straightforward application is the semantic communications \cite{grassucci2023generative,wu2024cddm,grassucci2024diffusion,10896580}, where the diffusion models are used to strengthen the noisy semantic.
Another important direction lies in the wireless channels} \cite{balevi2020high, MIMOdiffusion}, the authors investigated the MIMO channel estimation problem. They trained a diffusion model to generate MIMO channels and guided the generation process with the pilots and the received signals, which is an important reference for this work. Following this work, \cite{ma_diffusion_2024} investigated different diffusion methods for MIMO-OFDM scenarios, and \cite{fesl_diffusion-based_2024} showed that we can use the rough estimation result for initialization to accelerate the generation procedure without guidance. {Further, the channel reconstruction based on diffusion models can be extended to help sensing \cite{xu2025brownian}.}
All the above works do not take the data symbols into consideration, which greatly simplifies the problem, since the pilots are enough for a rough estimation and the diffusion model only needs to enhance a rough result.
{ Further, by introducing scenario information as conditions, the conditional diffusion model can work for various scenarios \cite{li2025wirelesschannelidentificationconditional, ni2025conditionaldiffusionmodeldrivengenerative}.}
However, in practical OFDM systems, not all subcarriers contain pilots, which means limited prior knowledge for those without pilots. Thus, in these settings, the guidance for some parts of the channel matrix becomes extremely poor, which brings extra difficulties.

Further considering the data symbols, \cite{zilberstein_joint_2024} proposed a diffusion-based unified channel and data generation algorithm for MIMO channels. { The authors of \cite{bhattacharya2025successive} also proposes to utilize data prior distributions in low-rank MIMO channels.} However, in this work, the data is distributed over time, thus the pilots are still sufficient for a rough channel estimation alone. Moreover, in \cite{zilberstein_joint_2024}, the data symbols are also generated by the diffusion model, the necessity of which we question. From our point of view, an independent uniform distribution over all possible symbols is precious and optimal to describe the data symbols transmitted through the channel. Given the channel, there are usually efficient signal estimation methods based on Bayesian inference, some of which are even proven to be optimal under some conditions. On the other hand, it is usually hard to capture the characteristics of a discrete distribution by NNs because of the drawbacks in gradient methods. Therefore, it is more desirable to retain some traditional data estimation units in the intelligent receivers.

{ Specific to the wireless channel estimation and signal recovery problem, most existing works focus on the MIMO system, which is similar to the discussed system \cite{balevi2020high, MIMOdiffusion,ma_diffusion_2024,fesl_diffusion-based_2024,xu2025brownian,zilberstein_joint_2024,li2025wirelesschannelidentificationconditional, ni2025conditionaldiffusionmodeldrivengenerative,bhattacharya2025successive}. However, in this paper, we focus on the OFDM system with a single UE. Unlike the MIMO estimation problem where we have a complete initial estimation, we do not have an initial rough estimation of the complete channel estimation. Instead, different parts of the channel estimation have completely different confidence levels, and we can obtain almost no prior estimation for the subcarriers without pilots. Thus, the OFDM receiver raises further requirements for the diffusion model. To conclude this section, we summarize the differences of the considered tasks and show the existing solutions in Table \ref{table::intro2}.}

\subsection{Conditional Diffusion Models}
Diffusion model is a popular generative AI framework \cite{ddpm, ddim}. They usually follow a simple gradual noise removal process for generation. In the original diffusion models, the generation process is quite random and inconclusive. That is, the framework only ensures that the output follows the data distribution indicated by the training dataset. However, in most applications, it is more desirable to generate under some provided conditions, i.e. conditional diffusion models \cite{Zhang_2023_ICCV, 11215776}.

To better guide the generation process of a conditional diffusion model, gradient-based guidance was introduced \cite{dhariwal2021diffusion}. Typically, we can use a scoring function (either explicit or expressed by another NN) to indicate how close the current generation is to the desired condition. Through gradient-based methods, we can then obtain the gradient of the score function with respect to the current generation result. Originally, we could use a separately trained classifier to guide the diffusion procedure towards the desired class \cite{dhariwal2021diffusion}. Later, this classifier was shown to be unnecessary with an appropriate sampling scheme \cite{ho2022classifierfreediffusionguidance}. Subsequent research also provided different scoring functions for other tasks, such as self-guided image editing \cite{epstein2023diffusion}, text-to-image generation with an additional language model \cite{zhang2023adding}, and even universal guidance, which provides a framework unifying different specific methods \cite{bansal2023universal}. To better understand gradient-based methods, a recent work also attempts to interpret them as an optimization problem \cite{guo2025gradient}.
\section{Problem Statement}
In this paper, we consider a simple OFDM uplink setting with a multi-antenna base station (BS) and a single-antenna user equipment (UE). We denote the channel as $\mathbf{H}\in\mathcal{C}^{N_\textrm{a}\times N_\textrm{c}}$, where $N_\textrm{a}$ and $N_\textrm{c}$ represent the number of antennas and subcarriers, respectively. We use $\bm{h}_i$ to indicate the $i$-th column of $\mathbf{H}$, i.e., the channel corresponding to the $i$-th subcarrier. The data is denoted by $\bm{x}\in\mathcal{C}^{N_\textrm{c}}$. Some subcarriers among all the $N_\textrm{c}$ subcarriers are selected as pilots, and others are data symbols. We denote the set of indices corresponding to pilots and data symbols as $\mathcal{P}$ and $\mathcal{D}$, respectively. The pilot symbols $x_i, i\in\mathcal{P}$ are all of unit power and known by the receiver. Meanwhile, data symbols $x_i, i\in\mathcal{D}$ are randomly and independently chosen from a symbol set $\mathcal{X}$, typically the symbol set of some quadrature amplitude modulation (QAM) modulation.
The received signal, denoted by $\mathbf{Y}$, can be expressed as
\begin{equation}
    \bm{y}_i = x_i \bm{h}_i + \bm{n}_i, \label{prob1}
\end{equation}
where $\bm{y}_i$ represents the $i$-th column of $\mathbf{Y}$ corresponding to the $i$-th subcarrier and $\bm{n}_i\sim \mathcal{C}\mathcal{N}(\bm{0}, \sigma_\textrm{n}^2\mathbf{I})$ is a complex Gaussian white noise. For the sake of simplicity, we use $\odot$ to describe column-wise multiplication such that \eqref{prob1} can also be written as
\begin{equation}
    \mathbf{Y} = \mathbf{H} \odot \bm{x} + \mathbf{N}, \label{prob2}
\end{equation}
{ where $\mathbf{N}$ is concatenated by all $\bm{n}_i$s.}
Thus, the task of the receiver can be formulated as recovering $\bm{x}$ through the received signal $\mathbf{Y}$ and an estimated $\sigma_\textrm{n}$. The prior knowledge includes the following. First, some elements in $\bm{x}$ are known as pilots, while the other elements can only take certain values. In addition, the channel $\mathbf{H}$ has some intrinsic characteristics determined by the physical process, which can be described by a dataset.
From the point of view of generative models, we can formulate the problem as trying to generate a tuple of $(\mathbf{H}, \bm{x})$ such that the above requirements are satisfied. Specifically, it is always easy to obtain a reasonable estimation of $\bm{x}$ that meets the corresponding requirements, whereas the structure of $\mathbf{H}$ is often difficult to explicitly depict. As in related works on image generation, we can always regard the output of a well-trained channel generation NN as reasonable estimations that fit the channel structural prior. The generation procedure can be terminated when we obtain a tuple $(\widehat{\mathbf{H}}, \widehat{\bm{x}})$ such that $\|\mathbf{Y}-\widehat{\mathbf{H}}\odot \widehat{\bm{x}}\|_\textrm{fro}^2<\xi^2
+\sigma^2_{\mr n}$, where $\|\cdot\|_\textrm{fro}$ indicates the Frobenius norm of a matrix and $\xi$ is a pre-given threshold value relative to the expected error, signal-to-noise ratios (SNR), and the selected modulation scheme.

\section{Proposed Framework}
\subsection{Framework Overview}
\begin{figure*}[h]
    \centering
    \includegraphics[width=0.9\linewidth]{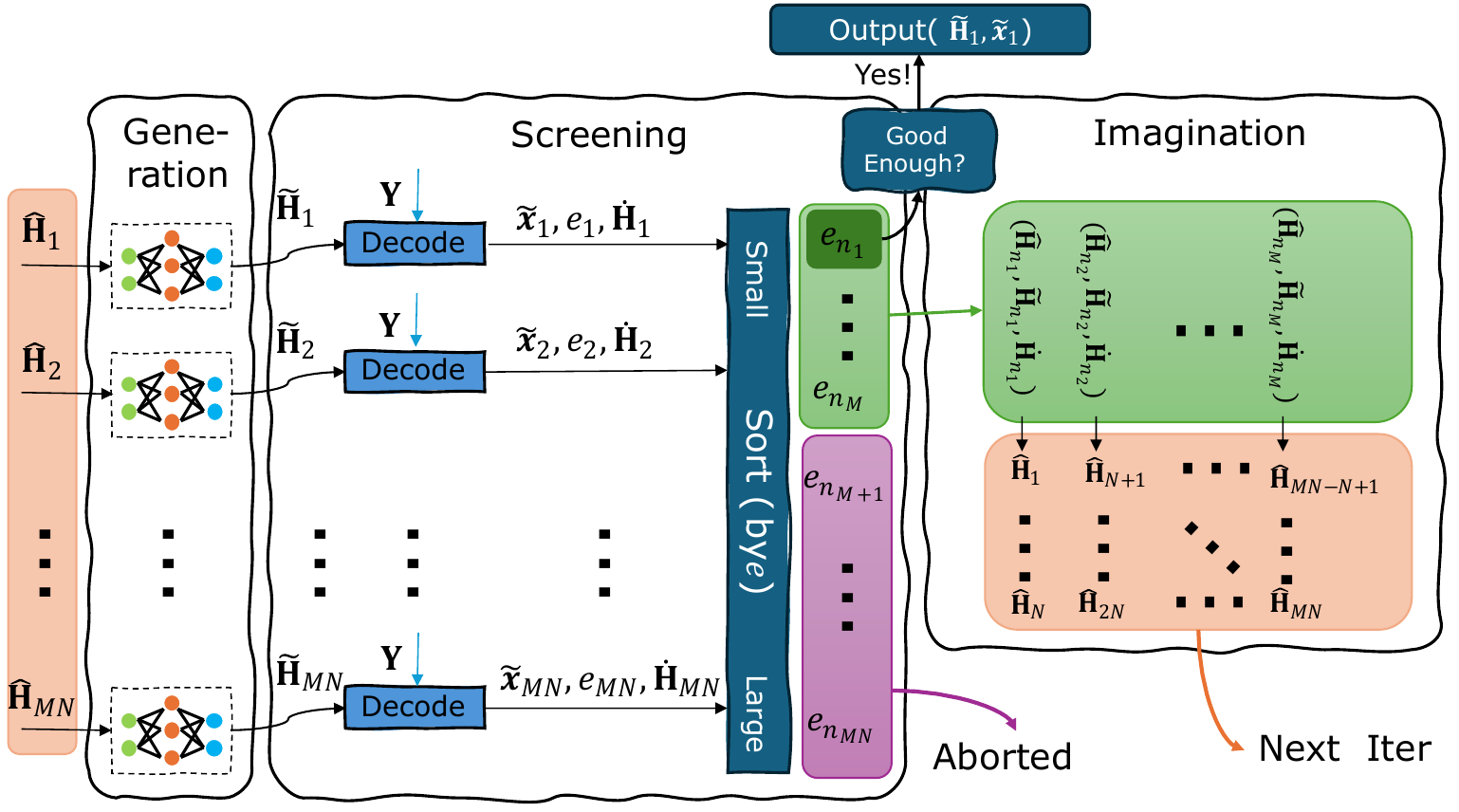}
    \caption{The workflow of the proposed diffusion-based receiver, only one iteration is shown.}
    \label{fig::workflow}
\end{figure*}
As mentioned above, we can use a channel generating NN based on the diffusion model with the hint of the received signal. The overall structure is illustrated in Fig. \ref{fig::workflow}. The forward diffusion follows the typical scheme where Gaussian white noise is gradually added to the channel $\mathbf{H}$, i.e.,
\begin{equation}
    \mathbf{H}^{(t+1)} = \sqrt{\alpha_t} \mathbf{H}^{(t)} + \sqrt{1-\alpha_t} \bm{\varepsilon}_t,
\end{equation}
where $\bm{\varepsilon}_t$ is a standard Gaussian white noise.

{To balance generation efficiency and stability, we adopt a hybrid approach between DDIM and DDPM as recommended in \cite{ddim}. While DDPM introduces stochasticity that enhances diversity, it may compromise initialization fidelity—especially problematic in our weakly guided setting. In contrast, DDIM offers deterministic sampling with fewer steps, which suits our task where imagination is less critical and computational efficiency is preferred. Additionally, training an autoencoder as required by latent diffusion models is nontrivial in our context, since the prior is an incomplete channel matrix rather than a full image or signal. Therefore, our chosen method provides a practical and effective compromise tailored to the problem structure.}

In backward denoising, each iteration starts with a group of rough estimations of the channel $\mathbf{H}$ at diffusion time $t$, denoted by $\{\widehat{\mathbf{H}}_1^{(t)}, \cdot, \widehat{\mathbf{H}}_{MN}^{(t)}\}$.
The NN is then used to generate a denoising result from each $\widehat{\mathbf{H}}_j^{(t)}$, denoted by $\widetilde{\mathbf{H}}_1^{(t_\textrm{next})}, \cdots, \widetilde{\mathbf{H}}_{MN}^{(t_\textrm{next})}$, which is called the ``generation'' step.
After that, we perform another step, namely the ``screening'' step, to eliminate redundant results. In the screening step, we use a traditional signal processing algorithm to obtain an estimation $\widetilde{\bm{x}}_j^{(t_\textrm{next})}$ of $\bm{x}$ independently for each of the $MN$ generated channels. We then sort the results by the $\|\mathbf{Y}-\widetilde{\mathbf{H}}_j^{(t_\textrm{next})}\odot\widetilde{\bm{x}}_j^{(t_\textrm{next})}\|_\textrm{fro}$ criterion. Some history estimations are also introduced here to ensure stability. If the best estimation fulfills the error threshold condition, it is output as the final estimation and the procedure is terminated. Otherwise, the best $M$ generated channels, denoted as $\{\widetilde{\mathbf{H}}_{n_1}^{(t_\textrm{next})}, \cdot, \widetilde{\mathbf{H}}_{n_M}^{(t_\textrm{next})}\}$, are retained and returned to the ``imagination'' unit for the next iteration, whereas others are aborted. The imagination unit then prepares the channel $\{\widehat{\mathbf{H}}_1^{(t)}, \cdot, \widehat{\mathbf{H}}_{MN}^{(t)}\}$. Specifically, $N$ random samples are generated from each $(\widehat{\mathbf{H}}_{n_j}^{(t)}, \widetilde{\mathbf{H}}_{n_j}^{(t_\textrm{next})})$ pair by the introduction of additive noise. In the remainder of this section, we dive into each part and provide the overall algorithm workflow.

\subsection{NN Training}\label{sec::training}
We mainly follow the typical DDIM training method \cite{ddim}. However, we conduct normalization on each channel matrix before inputting it into the NN. This is because the powers of wireless channels are usually distributed in an extremely wide range even under the same scenario due to the multipath effect. Since we need to coordinate the powers of the added noise and the input when training a diffusion model, such a large range of input power should be avoided if possible. We note that in the receiver, the power gain $\|\mathbf{H}\|_\textrm{fro}$ of the channel is always easy to obtain since the transmitted signal is always normalized. Thus, the receiver can conduct channel generation for channels with unit power and amplify them later.
Following \cite{ddim}, the loss function during training can be derived as follows.
\begin{align}
    \beta_t =& \beta_\textrm{max}t/T, \label{beta}\\
    \alpha_t =& 1-\beta_t, \label{alpha}\\
    \bar{\alpha_t} =& \prod_{\tau=1}^{t}\alpha_\tau \label{alpha2}\\
    \bar{\textbf{H}}_j =& \sqrt{\bar{\alpha}_{t_j}} \frac{\sqrt{N_\textrm{a}N_\textrm{c}}\textbf{H}_j}{\|\textbf{H}_j\|_\textrm{fro}} + \sqrt{1-\bar{\alpha}_{t_j}} \bm{\varepsilon}_j,\label{loss1} \\
    \ell =& \sum_j \|f(\bar{\textbf{H}}_j, t_j;\bm{\theta}) - \frac{\sqrt{N_\textrm{a}N_\textrm{c}}\textbf{H}_j}{\|\textbf{H}_j\|_\textrm{fro}}\|_\textrm{fro}^2,\label{loss2}
\end{align}
where $\beta_\textrm{max}$ and $T$ are given hyperparameters, $t_j$ is independently uniformly randomly drawn from $\{0,\cdots,T\}$, $\bm{\varepsilon}_j$ is a standard Gaussian white noise with the same shape as $\mathbf{H}_j$. $f(\mathbf{H}, t;\bm{\theta})$ denotes the NN that takes the channel matrix and the current time as inputs and $\bm{\theta}$ is its parameter set.
The training algorithm is also summarized in Alg. \ref{alg::train}.

\begin{algorithm}[t]
  \caption{Training the Channel Generation NN \label{alg::train}}
  \begin{algorithmic}
  \STATE \textbf{Input:} Initialized NN $f(\mathbf{H}, t;\bm{\theta})$, batch size $B$, learning 
  \STATE rate $\eta$, hyperparameters $\beta_\textrm{max}, T$, a channel dataset
  \STATE Calculate $\bar{\alpha}_t$ by \eqref{beta}, \eqref{alpha} and \eqref{alpha2};
  \WHILE{not converged}
  \STATE Sample $\{\textbf{H}_1,\cdots,\textbf{H}_B\}$ from the dataset;
  \STATE Uniformly sample $\{t_1,\cdots,t_B\}$ from $\{0,\cdots,T\}$;
  \STATE Independently sample $\{\bm{\varepsilon}_1,\cdot,\bm{\varepsilon}_B\}$ from standard Gaussian distribution with the same size as $\mathbf{H}$;
  \STATE Calculate loss $\ell$ by \eqref{loss1} and \eqref{loss2};
  \STATE $\bm{\theta}\leftarrow\bm{\theta}-\eta\partial\ell/\partial\bm{\theta}$;
  \ENDWHILE
  \end{algorithmic}
\end{algorithm}

\subsection{Receiver Algorithm by Diffusion}
\subsubsection{Initialization}
At the beginning of the receiver algorithm, we need to initialize the channel estimation alongside some necessary hyperparameters. Firstly, we need to obtain an original estimation of the channel gain $\sigma_H\triangleq\|\textbf{H}\|_\textrm{fro}/\sqrt{N_\textrm{a}N_\textrm{c}}$ since the NN is trained by ${\sqrt{N_\textrm{a}N_\textrm{c}}\textbf{H}}/\sigma_H$.
Given the fact that $\mathbb{E}(|\bm{x}_i|^2)=1$ always holds, we can easily obtain that
\begin{equation}
    \widehat{\sigma}_H = \|\textbf{Y}\|_\textrm{fro}/\sqrt{N_\textrm{a}N_\textrm{c}}.\label{sigma2}
\end{equation}


We can generate the original channel from the initial rough channel estimation. We can obtain a rough channel estimation of $\bm{h}_i$ by
\begin{equation}
    \widehat{\bm{h}}_i = \bm{y}_i/x_i, \quad \textrm{for} \quad i\in\mathcal{P},
\end{equation}
whereas we cannot obtain any prior estimation of $\bm{h}_i$ for $i\in\mathcal{D}$. To adapt this estimation into a reasonable range for the input of the diffusion model and add sufficient noise for generation stability, we can obtain the initial channel $\widehat{\mathbf{H}}_j^{(t_0)}$ by
\begin{align}
    \widehat{\bm{h}}_{j,i}^{(t_0)}=&\gamma \bm{y}_i/\sigma_H x_i + \delta\bm{\varepsilon}_{j,i}, \quad &\textrm{for} \quad i\in\mathcal{P},\label{init1}\\
    \widehat{\bm{h}}_{j,i}^{(t_0)}=&\delta\bm{\varepsilon}_{j,i}, \quad &\textrm{for} \quad i\in\mathcal{D},\label{init2}
\end{align}
where $\gamma,\delta\geq 0$ are pregiven values and all $\bm{\varepsilon}_{j,i}$s are independently sampled from the standard complex Gaussian distribution. We note that when $\gamma=0$ and $\delta=1$, it corresponds to a completely random initialization.

In addition to the initial estimation, we also need to determine the start time $t_0$ and the time series in the denoising process. We recall that in diffusion models the time step always corresponds to the current error. Thus, $t_0$, $\gamma$, and $\delta$ should follow the following relationships.
Firstly, we note that
\begin{equation}
    \bm{y}_i/x_i = \bm{h}_i + \bm{n}_i/x_i. \label{hatsigmaH}
\end{equation}
Thus, we have
\begin{equation}
    \widehat{\bm{h}}_{j,i}^{(t_0)} = (\gamma/\widehat{\sigma}_H)\bm{h}_i + (\gamma/\widehat{\sigma}_H x_i)\bm{n}_i + \delta\bm{\varepsilon}_{j,i}, \quad \textrm{for} \quad i\in\mathcal{P}.
\end{equation}
Since both $\bm{n}_i$ and $\bm{\varepsilon}_{j,i}$ are independent Gaussian variables and $|x_i|=1$, we have
\begin{equation}
    (\gamma/\widehat{\sigma}_H x_i)\bm{n}_i + \delta\bm{\varepsilon}_{j,i}\sim \mathcal{C}\mathcal{N}(\bm{0}, (\delta^2+\gamma^2\sigma_n^2/\widehat{\sigma}^2_H)\textbf{I}),\  \textrm{for}\ i\in\mathcal{P}.
\end{equation}
Thus, to be consistent with the forward diffusion process, we should have the energy constraint that $\|\widehat{\mathbf{H}}_{j}\|_\textrm{fro}^2=1$, which means
\begin{equation}    (\gamma^2+\delta^2+\gamma^2\sigma_n^2/\widehat{\sigma}^2_H)|\mathcal{P}|+\delta^2|\mathcal{D}|=|\mathcal{P}|+|\mathcal{D}|, 
\end{equation}i.e.,
\begin{align}
    0\leq\gamma\leq (|\mathcal{P}|+|\mathcal{D}|)/\sqrt{1+\sigma_n^2/\widehat{\sigma}^2_H}|\mathcal{P}|,\\
    \delta=\sqrt{1-\gamma^2(1+\sigma_n^2/\widehat{\sigma}^2_H)|\mathcal{P}|/(|\mathcal{P}|+|\mathcal{D}|)}.\label{delta}
\end{align}
{According to \eqref{beta}-\eqref{alpha2}, we can choose the initial time step by the energy proportion of the ground truth in the noisy original channel estimation.}
\begin{equation}
    t_0 = \arg\min_{t} \left|\bar{\alpha}_t-\frac{\gamma^2|\mathcal{P}|}{|\mathcal{P}|+|\mathcal{D}|}\right|,\label{t0}
\end{equation}
where $\bar{\alpha}_t$ is defined in \eqref{beta}-\eqref{alpha2}.
Given the total inference steps for generation $N_\textrm{gen}$, we can select the time series for diffusion evenly between $t_0$ and 0.

\subsubsection{Generation Step}
In the generation step, we directly use the NN output as the channel estimation, i.e.,
\begin{equation}
\widetilde{\mathbf{H}}_j^{(t_\textrm{next})}=f(\widehat{\mathbf{H}}_j^{(t)}, t;\bm{\theta}), \quad \textrm{for}\quad j = 1,\cdots,MN. \label{gen}
\end{equation}

\subsubsection{Memorized Screening Unit}
In the screening unit, we conduct a simple signal estimation algorithm to obtain $\widetilde{\bm{x}}_j^{(t_\textrm{next})}$ from $\textbf{Y}$ and each $\widetilde{\mathbf{H}}_j^{(t_\textrm{next})}$ independently. Specifically, we apply a hard decision for each symbol independently that
\begin{align}
    \widehat{x}_{j,i}^{(t_\textrm{next})} &= \widehat{\sigma}_H[\widetilde{\bm{h}}_{j,i}^{(t_\textrm{next})}]^\textrm{H}\bm{y}_i/[\widetilde{\bm{h}}_{j,i}^{(t_\textrm{next})}]^\textrm{H}\widetilde{\bm{h}}_{j,i}^{(t_\textrm{next})},\label{dec1}\\
    \widetilde{x}_{j,i}^{(t_\textrm{next})} &= x_i, \qquad\qquad\qquad\quad \textrm{for} \quad i \in \mathcal{P}, \label{dec2}\\
    \widetilde{x}_{j,i}^{(t_\textrm{next})} &= \textrm{Quantize}(\widehat{x}_{j,i}^{(t_\textrm{next})}), \quad \textrm{for} \quad i \in \mathcal{D}, \label{dec3}
\end{align}
where $\textrm{Quantize}(x)$ maps any input to the nearest constellation point in terms of Euclidean distance, which is a simple hard-decision demodulation method. We can then recover the expected received signal by the recovered channel and signal via
\begin{equation}
e_j^{(t_\textrm{next})}=\|\widehat{\sigma}_H\widetilde{\mathbf{H}}_j^{(t_\textrm{next})}\odot\widetilde{\bm{x}}_j^{(t_\textrm{next})}-\mathbf{Y}\|_\textrm{fro}^2.\label{err}
\end{equation}

We note that the channel generated in the later diffusion step is not necessarily better than the former ones. To restrict hallucination, i.e., to prevent the model from diffusing in an incorrect direction, we can apply a memory mechanism such that we substitute $\widetilde{\mathbf{H}}_j^{(t_\textrm{next})}$ by $\widetilde{\mathbf{H}}_j^{(t)}$ if $e_j^{(t)}<e_j^{(t_\textrm{next})}$.
After that, by sorting all $e_j^{(t_\textrm{next})}$s in ascending order, we have $\{n_1,\cdots,n_{MN}\}$ such that $e_{n_1}^{(t_\textrm{next})}\leq\cdots\leq e_{n_{MN}}^{(t_\textrm{next})}$. If the smallest error has reached the provided error gate $\xi^2+\sigma_{\mr n}^2$ or the diffusion procedure has finished, the algorithm is terminated and the corresponding estimation becomes the output. That is, if $e_{n_1}^{(t_\textrm{next})}\leq\xi^2+\sigma_{\mr n}^2$ or $t_\textrm{next}\leq 0$, we adopt $\widehat{\sigma}_H\widetilde{\mathbf{H}}_{n_1}^{(t_\textrm{next})}$ and $\widetilde{\bm{x}}_{n_1}^{(t_\textrm{next})}$ as the final channel and data estimation, respectively. Otherwise, $\{\widetilde{\mathbf{H}}_{n_1}^{(t_\textrm{next})}, \cdots, \widetilde{\mathbf{H}}_{n_M}^{(t_\textrm{next})}\}$ are passed to the imagination unit to continue the overall iteration.

\subsubsection{Imagination Unit}
{In the imagination unit, we still follow the diffusion procedure, which is a direct combination of DDPM and DDIM \cite{ddim}. Such method is used since DDPM
}

Specifically, we have
\begin{align}
\widetilde{\bm{\varepsilon}}_{n_j}^{(t_\textrm{next})}=&(\widehat{\mathbf{H}}_{n_j}^{(t)}-\sqrt{\bar{\alpha}_t}\widetilde{\mathbf{H}}_{n_j}^{(t_\textrm{next})})/\sqrt{1-\bar{\alpha}_t},\label{epsilon}\\
\sigma_t =& \sqrt{\frac{1-\bar{\alpha}_{t_\textrm{next}}}{1-\bar{\alpha}_t}}\cdot\sqrt{1-\frac{\bar{\alpha}_t}{\bar{\alpha}_{t_\textrm{next}}}},\label{sigma}\\
\widehat{\mathbf{H}}_{(j-1)N+i}^{(t_\textrm{next})}=&\sqrt{\bar{\alpha}_{t_\textrm{next}}}\widetilde{\mathbf{H}}_{n_j}^{(t_\textrm{next})}+ \zeta\sigma_t\bm{\varepsilon}_{j, i}\nonumber\\
&  + \sqrt{1-\bar{\alpha}_{t_\textrm{next}}-\zeta\sigma_t^2}\widetilde{\bm{\varepsilon}}_{n_j}^{(t_\textrm{next})} .\label{imag}
\end{align}
In \eqref{epsilon}, we estimate the error in $\widehat{\mathbf{H}}_{n_j}^{(t)}$, that is, trying to recover $\bm{\varepsilon}$ in \eqref{loss1}. $\sigma_t$ provides the weight of the added noise (imagination), which is determined by a hyperparameter $\zeta$. $\bm{\varepsilon}_{j, i}$ is independently drawn from a standard complex Gaussian distribution, providing several parallel candidates for the next iteration of channel generation.

\begin{algorithm}[t]
  \caption{Diffusion-Based OFDM Receiver \label{alg::diff}}
  \begin{algorithmic}
  \STATE \textbf{Input:} Well-trained NN $f(\mathbf{H}, t;\bm{\theta})$, received signal $\mathbf{Y}$, 
  \STATE noise power $\sigma^2_n$, hyperparameters $\gamma$, $\zeta$, $\xi$, and $N_\textrm{gen}$;
  \STATE Estimate $\widehat{\sigma}_H$ by \eqref{sigma2};
  \STATE Calculate $\delta$ and $t_0$ by \eqref{delta} and \eqref{t0};
  \STATE Initialize $MN$ channel matrices $\widehat{\mathbf{H}}_j^{(t_0)}$ by \eqref{init1} and \eqref{init2};
  \WHILE{not terminated}
  \STATE $t_\textrm{next} \leftarrow t-\textrm{round}(t_0/N_\textrm{gen})$;
  \STATE Generate $MN$ channel matrices $\widetilde{\mathbf{H}}_j^{(t_\textrm{next})}$ by \eqref{gen};
  \STATE Estimate $\widetilde{\bm{x}}_j^{(t_\textrm{next})}$ by \eqref{dec1}-\eqref{dec3};
  \STATE Calculate the errors $e_j^{(t_\textrm{next})}$ by \eqref{err};
  \IF{$e_j^{(t_\textrm{next})}>e_j^{(t)}$}
  \STATE $\widetilde{\mathbf{H}}_j^{(t_\textrm{next})}\leftarrow\widetilde{\mathbf{H}}_j^{(t)}$;\quad $\widetilde{\bm{x}}_j^{(t_\textrm{next})}\leftarrow\widetilde{\bm{x}}_j^{(t)}$;
  \ENDIF
  \STATE Sort all $e_j^{(t_\textrm{next})}$s by ascending order and get the 
  \STATE indices $\{n_1,\cdots,n_{MN}\}$ of the sorted sequence;
  \IF{$t_\textrm{next}\leq 0$ or $e_{n_1}^{(t_\textrm{next})}\leq\xi^2+\sigma_{\mr n}^2$}
  \STATE Output $\widehat{\sigma}_H\widetilde{\mathbf{H}}_{n_1}^{(t_\textrm{next})}$ and $\widetilde{\bm{x}}_{n_1}^{(t_\textrm{next})}$ as the estimation;
  \STATE Terminate the algorithm;
  \ENDIF
  \STATE Keep $\widetilde{\mathbf{H}}_{n_1}^{(t_\textrm{next})},\cdots,\widetilde{\mathbf{H}}_{n_M}^{(t_\textrm{next})}$ and abort others;
  \STATE Generate $MN$ channel matrices $\widehat{\mathbf{H}}_j^{(t_\textrm{next})}$ by \eqref{epsilon}-\eqref{imag};
  \STATE $t\leftarrow t_\textrm{next}$;
  \ENDWHILE
  \end{algorithmic}
\end{algorithm}

\subsubsection{Overall Receiver Algorithm}
In conclusion, the proposed diffusion-based receiver algorithm is summarized in Alg. \ref{alg::diff}.

\subsubsection{Theoretical Discussion on the Screening Method}{
Our proposed iterative diffusion and screening process can be viewed as a simplified form of classical sampling-based Bayesian inference techniques. Specifically, by repeatedly sampling from the prior distribution and applying a filtering mechanism based on the conditional likelihood, the retained samples progressively approximate the posterior distribution. This approach is conceptually related to importance sampling and rejection sampling, where samples from the prior are weighted or selected according to their likelihood under the observed data \cite{doucet2001sequential, skilling2006nested}.

While our method does not explicitly compute posterior weights or rely on complex resampling schemes, it leverages the same principle that high-likelihood regions of the prior are preferentially retained, thereby concentrating the sample distribution toward the posterior. This strategy has been shown to be effective in various approximate inference frameworks, including particle filtering and nested sampling \cite{arulampalam2002tutorial, skilling2006nested}. Our design simplifies these mechanisms to suit the structure of our model and task, trading theoretical guarantees for computational efficiency and empirical robustness.}

\subsection{Advantages and Disadvantages of the Diffusion-Based Receiver}
Here, we briefly discuss the advantages and disadvantages of the proposed diffusion-based OFDM receiver compared to traditional and AI-based methods.

\subsubsection{Comprehending the Channel Structure}
Compared to conventional approaches, NNs offer enhanced capabilities in depicting the inherent characteristics of channels. Given the sparse nature of the paths within the spatial domain, minimal channel observations are theoretically adequate for accurate recovery. Additionally, the presence of multiple antennas allows the exploitation of antenna domain data in characterization. However, conventional channel modeling and interpolation techniques often face a performance drop under noisy input conditions. Consequently, the use of NNs enables a substantial reduction in pilot overhead without significantly compromising the signal recovery accuracy.

\subsubsection{Generalization to Different Transmission Schemes}
It is important to note that during the training phase of the diffusion model, specific pilot schemes and modulation methods are not required. Consequently, once properly trained, the NN has the flexibility to adapt to various pilot and modulation configurations. This adaptability allows modifications to the transmission scheme as channel conditions fluctuate, which presents a significant benefit. Specifically, as the channel state transitions, the NN model allows for a shift to an alternative transmission configuration characterized by varying pilot densities and symbol modulation techniques without necessitating a retraining process.

{ Further, we note that the proposed scheme does not necessarily rely on any specific NN topology or signal processing method. The backbone NN is only required to be powerful enough for channel characterization in the considered scenario, and the signal processing method is only used to judge different channel estimation candidates.
Thus, with other more practical and complicated scenarios with different transmission schemes, we can turn to more efficient backbone NN topologies such as those in \cite{ju2024transformerassistedparametriccsifeedback,zhang2025residualcrossattentiontransformerbasedmultiuser}. 
With practical resource block structures, we can directly substitute the signal processing module by more powerful ones.
With multiple UEs, we can conduct the diffusion model in parallel corresponding to each UE, and use multi-user signal processing algorithms in the screening step.}

\subsubsection{Utilizing the Sparsity of Signal}
One key feature of the communication system lies in the sparsity of transmitted signals. In all digital communication systems, the sparsity introduced by constellation diagrams should be carefully considered for signal estimation. However, NNs are not well-suited to directly deal with such discrete values. Although there is a wealth of work on quantized NNs \cite{gholami2022survey}, the target is always to approximate the performance of full-precision ones instead of utilizing the discreteness.
However, with the screening procedures, the denoising process can indirectly utilize such information through traditional signal processing units, providing additional assistance to the channel generation procedure.

\subsubsection{Complexity and Latency Concerns}
On the other hand, the introduction of NNs unavoidably brings high complexity, which may result in prohibitive latency. Since NN only consists of several small fully-connected layers, it usually does not bring too much computational complexity \cite{mixer} during deployment, which is in the same order as matrix multiplication. However, in the denoising process, the basic NN unit is deployed for each potential channel estimation in each diffusion step. In summary, we need to run $MNN_\mr{gen}$ times of the NN unit and the traditional signal processing module if the algorithm is not terminated before the last step. Thus, the choice of hyperparameters becomes extremely limited and the potential of the proposed method cannot be fully exploited in reality due to the complexity requirements.

\subsubsection{Hallucination}
Although we can use the screening and memory mechanisms to restrict hallucination in the proposed scheme, it is still a potential drawback in the diffusion scheme. Under bad channel conditions, there may be multiple pairs $(\widehat{\textbf{H}}, \bm{x})$ leading to a similar value of $\|\textbf{H}-\widehat{\textbf{Y}}\odot\bm{x}\|_\textrm{fro}$, resulting in unavoidable ambiguity. Especially if the system starts to diffuse in an incorrect direction, it is almost impossible to correct it in the proposed scheme. In such scenarios, although the proposed scheme is very likely to perform well, there is a small possibility of complete failure even under some good conditions.

\section{Numerical Results}
\subsection{NN Structure and Dataset}
\begin{figure}[h]
    \centering
    \includegraphics[width=0.9\linewidth]{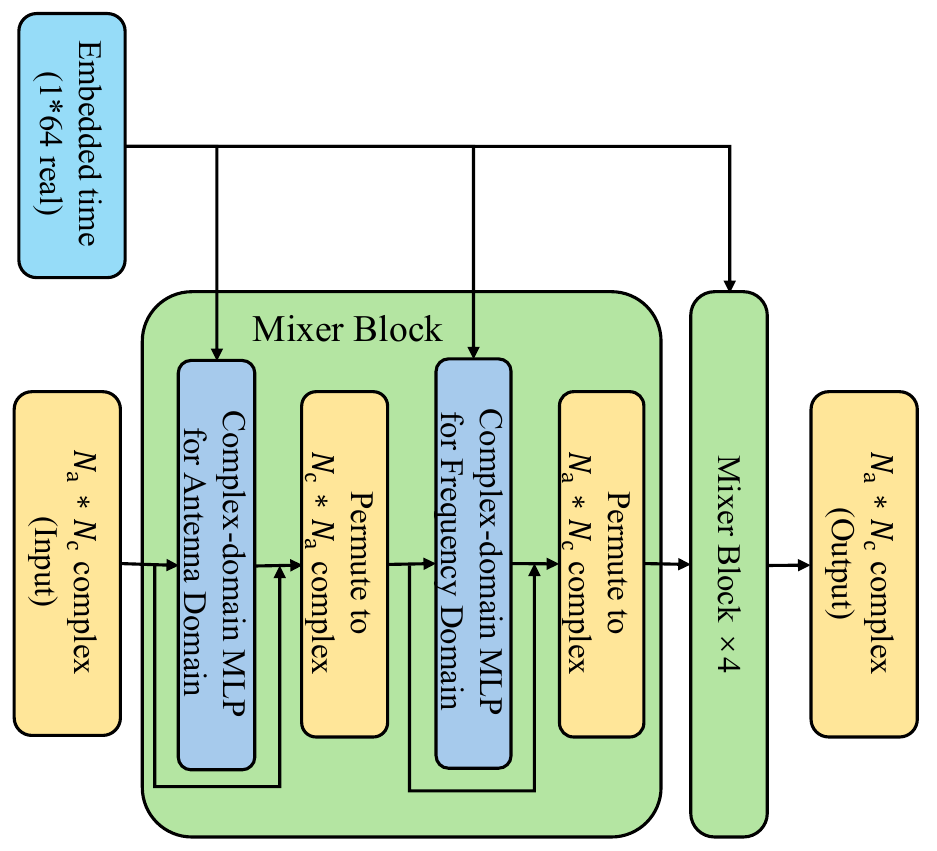}
    \vspace{0.5 cm}
    \includegraphics[width=0.9\linewidth]{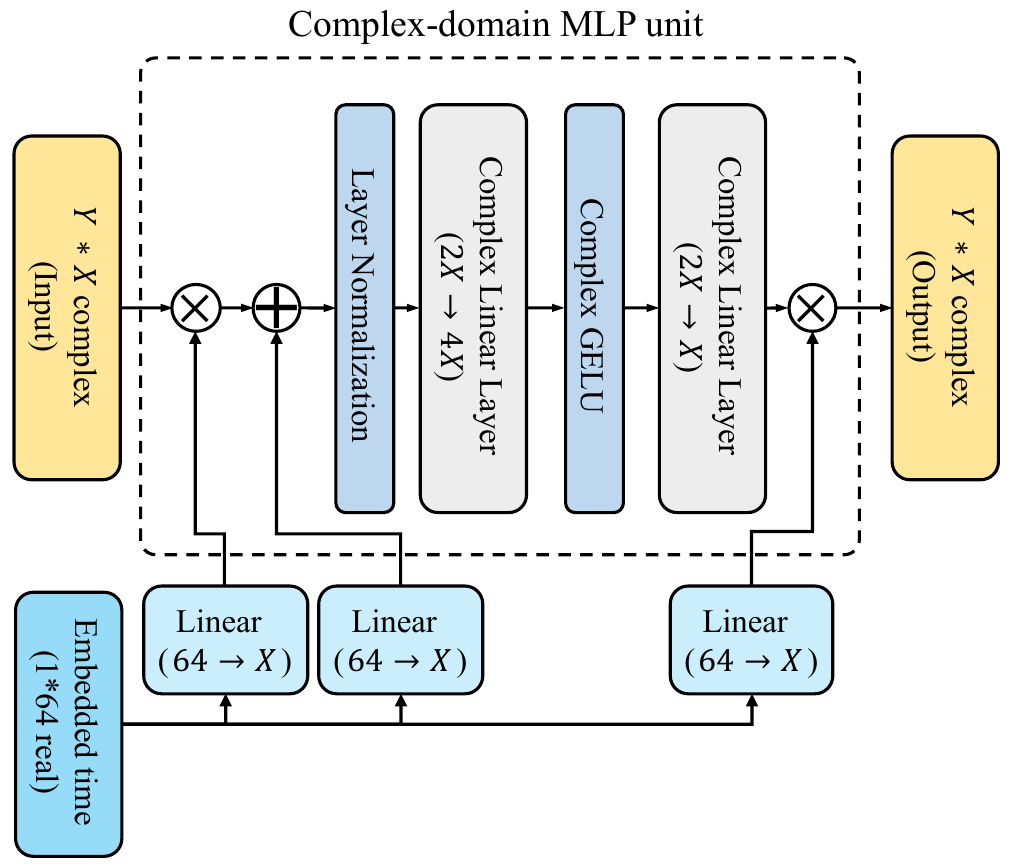}
    \caption{The NN used in this paper.}
    \label{fig::NN}
\end{figure}
In this paper, we mainly follow the experimental setting in \cite{mixer}. Specifically, we use the NN as illustrated in Fig. \ref{fig::NN}. The NN structure mainly follows the MLP-Mixer framework as in \cite{mixer}, which has been proven to be efficient in characterizing MIMO-OFDM channels. It is a lightweight NN that only costs $\sim$428k parameters and $\sim$33.2 MFLOPs calculation overhead for each forward computation of a $N_\mathrm{a}=32, N_\mathrm{p}=64$ channel matrix. We also added the time embedding unit to the NN for the diffusion procedure.

{ To ensure a fair comparison between diffusion and non-diffusion schemes, we adopt the same backbone NN architecture for both. While many alternative architectures exist with varying complexity and performance, the focus of this work is on the diffusion mechanism itself rather than optimizing the backbone design. Therefore, we select a lightweight yet effective model to balance performance and computational cost. Notably, some recent models \cite{ju2024transformerassistedparametriccsifeedback,zhang2025residualcrossattentiontransformerbasedmultiuser} tend to be significantly deeper, and in practical deployment, linear layers offer faster inference than attention-based counterparts.}


We use the DeepMIMO \cite{deepmimo} dataset to verify the proposed method. We use the ``O1'' outdoor scenario,
where each data corresponds to the channel between a point in the 36 m $\times$ 59.8 m area and a BS beside the road. The detailed area and training settings can be found in our online source code\footnote{https://github.com/yuzhiyang123/channel-diffusion}. UEs are uniformly distributed with a spacing of 20 cm. There are a total of 54,300 channel matrices in this area, 80\% of which are randomly drawn for the training dataset and the remaining are used as the testing dataset.
The frequency of the first subcarrier is 3.5 GHz, and there are a total of $N_\textrm{c}=64$ subcarriers with an interval of 300 kHz. The UE is equipped with one antenna, and the BS has $N_\textrm{a}=32$ unified linear array antennas with spacing of half wavelength, that is, 4.3 cm.
$N_\textrm{p}$ of the total $N_\textrm{c}$ subcarriers are chosen evenly with the same spacing as the pilots.
The symbols in the remaining subcarriers are randomly selected from the QPSK or the 16-QAM constellation points.


\subsection{Training}
\begin{figure}[h]
    \centering
    \includegraphics[width=1\linewidth]{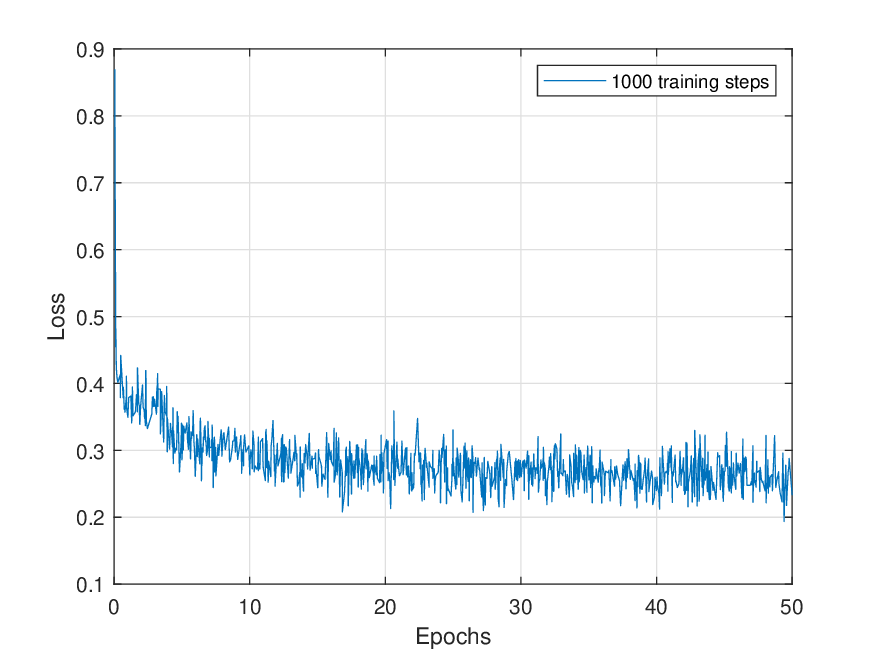}
    \caption{The loss vs epoch curve during training.}
    \label{fig::training}
\end{figure}
We train our NN with $T=1000$ time steps, following the training method introduced in Section \ref{sec::training}. The convergence curve shown in Fig. \ref{fig::training} proves the effectiveness of the NN structure. We note that since the noise levels are randomly chosen, the fluctuations in the curves do not mean instability. In the following part of this paper, we use this trained model for all generation tasks, except for certain comparison methods, which also demonstrate the advantage of model universality in the diffusion-based method.

\subsection{Basic Channel Generation Settings}
\begin{table}[h]
    \centering
    \begin{tabular}{|c|c|c|}
        \hline
        \textbf{Parameter} & \textbf{Meaning / Usage}& \textbf{Value} \\
        \hline
        $N_{\mr P}$ & \# of pilots & 4 (4.8 MHz spacing) \\
        $\beta_\mr{max}$ & Training hyperparam. & 0.2 \\
        $\gamma$ & Initialization strength & 14.4 \\
        $\zeta$ & Imagination proportion & 0.4 \\
        $M$ & \# of parallel imagination & 16 \\
        $N$ & \# of remaining after screening & 8 \\
        $\xi$ & Early quit gate & 0\\
        $N_\mr{gen}$ & Max generation steps & 100 \\
        SNR & SNR & 0 dB\\
        \hline
    \end{tabular}
    \caption{List of Default Parameters and Meanings}
    \label{table::setting}
\end{table}
In the following sections, we show the results of channel generation. Unless otherwise stated, we use the data in the test dataset and the default settings in Table \ref{table::setting}. Without loss of generality, we set $x_i=1$ for all $i\in\mathcal{P}$ in the simulation since they are always independently processed.
We note that we do not activate the early quit mechanism to show the full diffusion procedure by default, i.e. $\xi=0$, while investigating its influence by showing the generation. We also note that $\gamma|\mathcal{P}|/(|\mathcal{P}|+|\mathcal{D}|)$ indicates the power proportion of the estimation in the initialization. In the default setting, it is 0.9, and when the pilot scheme changes, we also adapt $\gamma$ accordingly to maintain the proportion.

{\begin{table}[h]
\begin{tabular}{|l|l|l|l|}\hline
time / comp                                                                      & $M=16,N=8$ & $M=N=4$ & $M=N=1$ \\ \hline
\begin{tabular}[c]{@{}l@{}}$N_\text{gen}=20$,\\ batch size 1\end{tabular}  & 448.7 / 85.0     & 455.3 / 10.6  & 450.2 / 0.66   \\\hline
\begin{tabular}[c]{@{}l@{}}$N_\text{gen}=100$,\\ batch size 1\end{tabular} & 2539.2 / 425    & 2519.3 / 53.1 & 2470.1 / 3.3 \\\hline
\begin{tabular}[c]{@{}l@{}}$N_\text{gen}=20$,\\ batch size 8\end{tabular}  & 756.1 / 679.9     & 460.4 / 85.0   & 449.0 / 5.3  \\\hline
\end{tabular}
\caption{Operation time and computational cost of different settings. The results are in (ms/batch / GFLOPs) and averaged over 10 batches.}\label{tab:time}
\end{table}}
{To provide a more comprehensive understanding of the computational burden associated with our proposed method, we conducted empirical latency measurements using an NVIDIA RTX A6000 GPU with 48GB memory. The results, summarized in Table \ref{tab:time}, show that although the theoretical complexity scales as $M \times N \times N_{\text{gen}}$, the actual inference time does not increase proportionally. This is primarily due to the parallel computation capabilities of modern GPUs, which effectively offset the added complexity introduced by the imagination-screening structure.

In the single-user setting considered in this work, the batch size is set to 1 due to the sequential nature of signal generation. For multi-user extensions, the batch size scales with the number of users, further leveraging GPU parallelism. As shown in Table \ref{tab:time}, the additional computational complexity does not translate into a significant increase in latency, demonstrating the practical feasibility of our approach under current hardware conditions.

Nevertheless, we acknowledge that the observed inference latency remains higher than the stringent requirements of URLLC systems. This is partly due to the use of general-purpose hardware and the lack of runtime-specific code optimization. To address this, we note that recent advances in diffusion model acceleration offer promising directions for reducing inference time. Techniques such as quantization \cite{zeng2025diffusion}, feature caching \cite{liu2025smoothcache, liu2025survey}, encoder skipping and reuse \cite{li2024faster}, and distributed parallel inference \cite{li2024distrifusion} have demonstrated substantial speedups—ranging from 5× to over 30×—with minimal degradation in generation quality. These methods are training-free and architecture-agnostic, making them suitable for integration into our framework.}

\begin{table}[b]
    \centering
    \begin{tabular}{c|l|ccc}
        \hline
        & Algorithm & SNR 0 dB & SNR -2 dB & SNR -4 dB \\
        \hline
        \multirow{5}{*}{\rotatebox{90}{$N_\mr{P}=4$}}&Diffusion (16 QAM) & 0.13 & 0.14 & 0.22 \\
        &DiffTraining & 0.13 & 0.17 & 0.27 \\
        &Direct-Diff & 0.22 & 0.77 & 2.12 \\
        &Direct-Gaussian & 0.14 & 0.36 & 0.90 \\
        &Direct-Noiseless & 2.39 & 5.28 & 12.22 \\
        \hline
        \multirow{5}{*}{\rotatebox{90}{$N_\mr{P}=6$}}&Diffusion (16 QAM) & 0.04 & 0.05 & 0.10 \\
        &DiffTraining & 0.06 & 0.07 & 0.12 \\
        &Direct-Diff & 0.20 & 0.79 & 2.10 \\
        &Direct-Gaussian & 0.15 & 0.34 & 0.78 \\
        &Direct-Noiseless & 1.14 & 2.83 & 7.01 \\
        \hline
    \end{tabular}
    \caption{Performance of different algorithms under various SNRs}
    \label{table::comparison}
\end{table}

\begin{figure*}[t]
    \centering
    \includegraphics[width=0.78\linewidth]{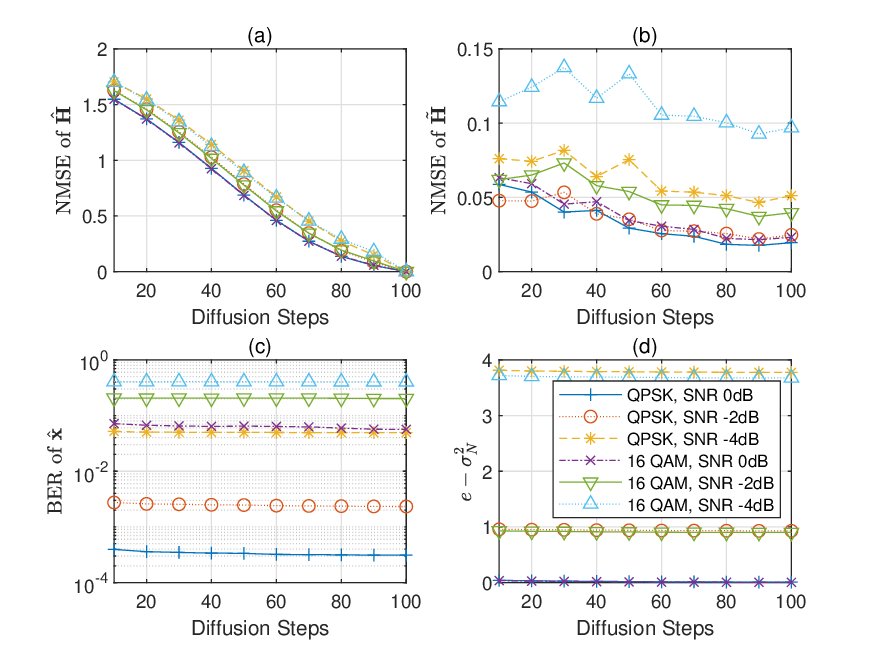}
    \caption{Results under different SNRs and modulations.}
    \label{fig::SNR}
\end{figure*}
\subsection{Comparison with Non-Diffusion Models}
Here, the total number of pilots $N_\textrm{p}$ is set to 4 or 6, and we use the following benchmark training methods to compare the proposed method with traditional NN methods without diffusion. We consider the following four methods as benchmarks. The first is ``DiffTraining'', which uses exactly the same model as the proposed diffusion-based method. However, instead of the diffusion procedure, we directly use its output as the final channel prediction, which can be regarded as a one-step diffusion. The other three methods use NNs without time embedding that receive only partial channels as input. In these methods, the NN is exactly the same as in \cite{mixer} and is specific to the selected pilot scheme, namely ``Direct'' methods. The difference lies in the training methods. In the ``Diff'' method, we use exactly the same noise pattern as in the diffusion-based method although it is not used as a diffusion model. In the ``Gaussian'' method, we use a simple noise-adding method that we first randomly choose a uniformly distributed noise power $\alpha\sim U(0, \|\mathbf{H}\|_2^2)$. Then, we use channel $\mathbf{H}_{\mr P}+\sqrt\alpha \bm{n}$ instead of $\mathbf{H}_{\mr P}$ as the NN input during training. Finally, in the ``Noiseless'' method, we directly use the original sliced channel matrices as the NN input, corresponding to the original channel mapping task.

The results under different SNRs and pilot schemes are shown in Table \ref{table::comparison}. We find that the diffusion-based method with appropriate time embedding and channel estimation outperforms in channel estimation. Moreover, we observe another important advantage of the diffusion model: it does not require specific training regarding the pilot schemes. Rather than retraining a NN for each noise condition and pilot scheme, the diffusion model enables the NN to be reused across various conditions.

\subsection{Basic Channel Generation Results}
Here, we show the evolution of the following four variables during the diffusion procedure, where the total number of pilots $N_\textrm{p}=4$. $\widehat{\mathbf{H}}$ is the main diffusion variable, which approaches the true channel over time. The error of $\widetilde{\bm x}$ is the key performance indicator of the receiver.
$\widetilde{\mathbf{H}}$ represents the best channel generation we obtain in the current step, and its error is positively related to that of $\widetilde{\bm x}$, becoming especially intuitive when SNR is high. Also, since $\widetilde{\mathbf{H}}$ is directly used for the final output, the diffusion model works well only when $\|\widetilde{\mathbf{H}}-\mathbf{H}\|_2^2$ gradually decreases. Thus, the diffusion should stop when $|\widetilde{\mathbf{H}} - \mathbf{H}|_2^2$ stops decreasing.
Finally, we also show the error $e$ as the error metric used to determine termination.

\begin{figure}[h]
    \centering
    \includegraphics[width=0.9\linewidth]{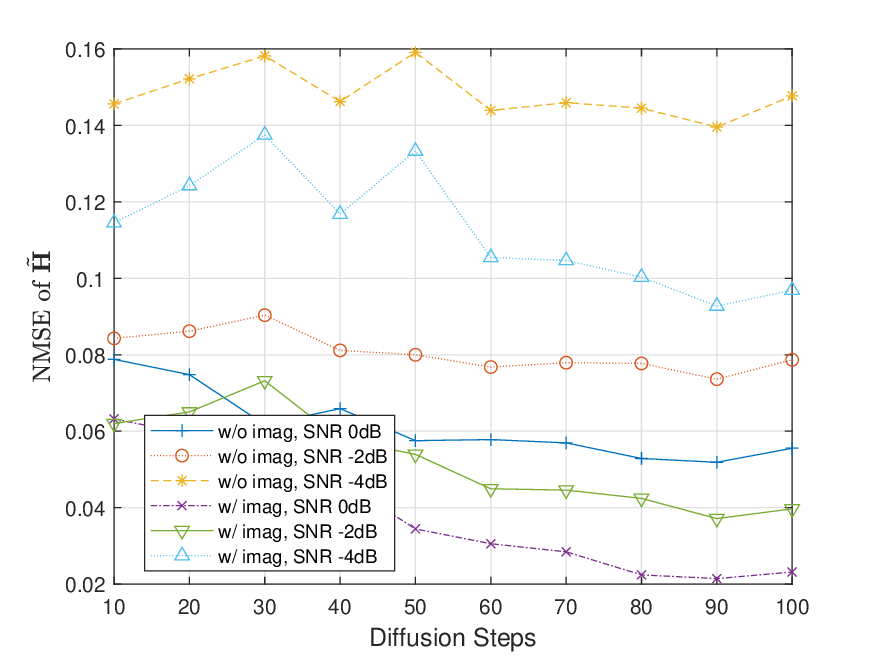}
    \caption{Channel generation results with and without enabling imagination module.}
    \label{fig::imag}
\end{figure}
In Fig. \ref{fig::SNR}, we show the performance under different SNRs and modulation schemes. We note that the considered system contains $N_\mr{ant}=32$ receiving antennas for each symbol; thus, the required SNR for successful transmission becomes relatively low.
From Fig. \ref{fig::SNR}, we find that the proposed algorithm works well and can generate better channels as the reverse diffusion progresses. From the curves of $\widetilde{\mathbf{H}}$, we observe that QPSK outperforms 16-QAM at the same noise level, since the sparser constellation diagram can provide more information. From the curves of $\widetilde{\bm x}$, we see a performance gain during the reverse diffusion procedure in terms of recovery accuracy, which appears to be more remarkable when the channel condition is good.
Fig. \ref{fig::imag} shows the effect of the imagination part ($\zeta=0.4$ or 0) on the channel generation result, which demonstrates the effectiveness of our imagination component. { Besides the results here, we have also conducted experiments with other straightforward methods that fails to perform well, which is shown in Appendix \ref{app}.}

\subsection{Denoising Procedure}

\begin{figure*}[t]
    \centering
    \includegraphics[width=0.78\linewidth]{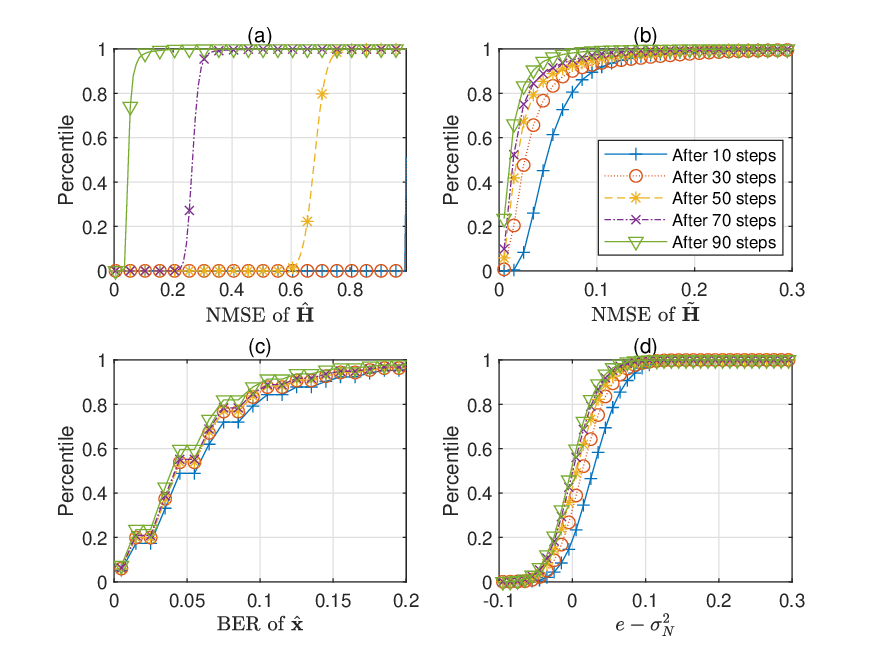}
    \caption{A typical diffusion procedure under SNR 0 dB.}
    \label{fig::PDF}
\end{figure*}
To better show the diffusion procedure and the impact of the early stop mechanism, we also show the PDF of the errors in Fig. \ref{fig::PDF}, which illustrates the evolution of the error distribution. We observe that most of the data reach a satisfactory channel estimation accuracy after 30 steps of generation. The average error is largely determined by a few bad results, and the performance gain through diffusion procedure is mainly provided by them. Thus, with an appropriate early stop gate, the diffusion procedure is likely to terminate early without significantly degrading performance. Since the algorithm complexity is relatively high, it is important to determine this gate according to the desired error to obtain a balance between complexity and performance.
We also note that the recovery error $e$ is sometimes less than the noise power $\sigma_N^2$. This does not mean that the performance exceeds physical limitations. Instead, it is probably brought by some accidental error, indicating the error criterion may be invalid and also proving the necessity of the early stop mechanism.
\subsection{Performance Under Different Screening Size}

\begin{figure}[h]
    \centering
    \includegraphics[width=0.9\linewidth]{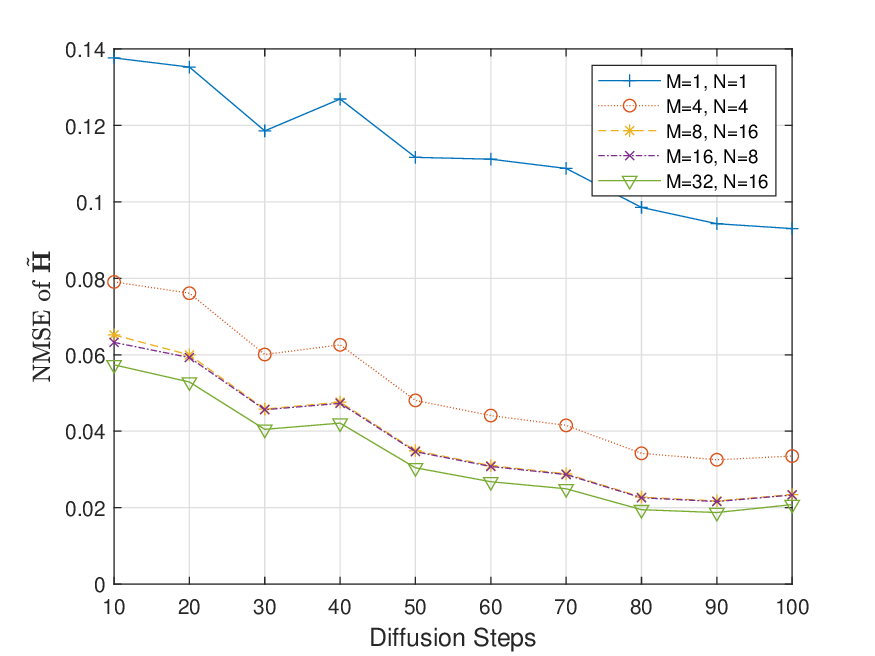}
    \caption{Channel generation results with different screening parameters.}
    \label{fig::screening}
\end{figure}
The parameters $M$ and $N$ in the screening unit determine the overall complexity. Here, we show the results with different $(M,N)$ pairs to show the performance of the screening unit in Fig. \ref{fig::screening}. We can easily find that the larger $M$ and $N$ we set, the better the result, which is intuitive. However, as their values increase, the complexity of the overall algorithm also increases dramatically in terms of $\mathcal{O}(MN)$. Therefore, we should find an appropriate point for the complexity-performance tradeoff.

\subsection{Results Under Different Imagination Levels}

\begin{figure}[h]
    \centering
    \includegraphics[width=0.9\linewidth]{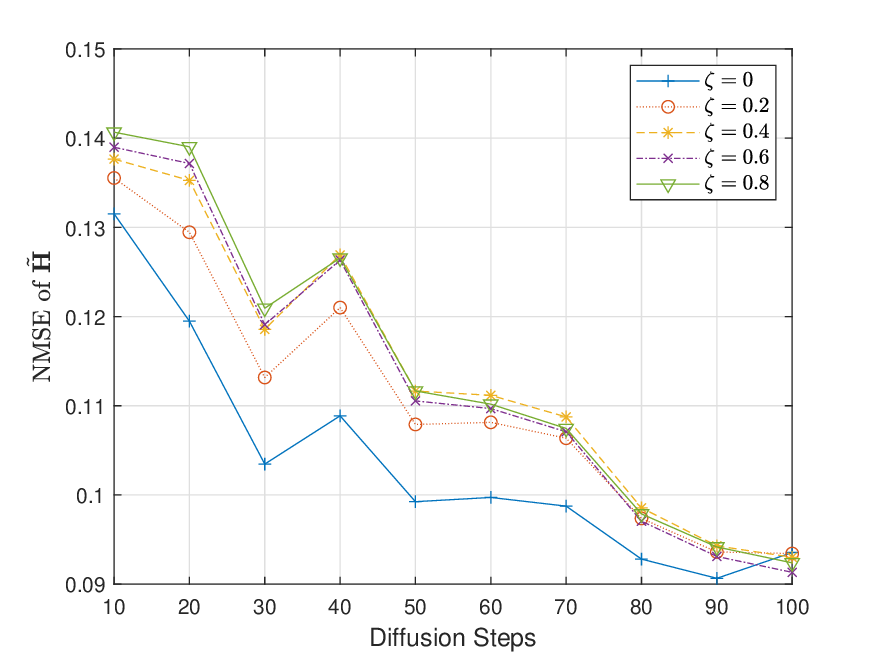}
    \caption{Channel generation results with different $\zeta$ and $M=N=1$.}
    \label{fig::eta1}
\end{figure}

\begin{figure}[h]
    \centering
    \includegraphics[width=0.9\linewidth]{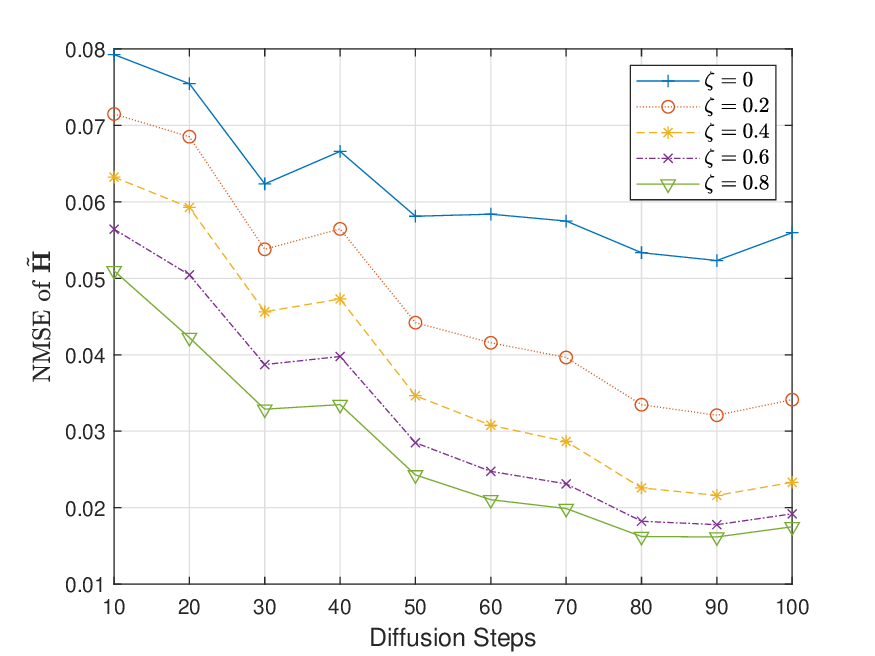}
    \caption{Channel generation results with different $\zeta$ and $M=16$, $N=8$.}
    \label{fig::eta2}
\end{figure}
Figs. \ref{fig::eta1} and \ref{fig::eta2} show the results corresponding to different $\zeta$ values and different screening parameters. A very interesting finding is that there are contradictory trends indicated by both figures. A possible explanation for this interesting phenomenon is that random directions usually lead to a worse result, but the best ones in a group of independent samplings usually lead to a better result, which is quite similar to evolution. Therefore, when we do not actually perform the screening, imagination is harmful, whereas if we perform appropriate screening on a large group, imagination becomes a key to improving performance.

\subsection{Results Under Different Initializations}
\begin{figure}[h]
    \centering
    \includegraphics[width=0.9\linewidth]{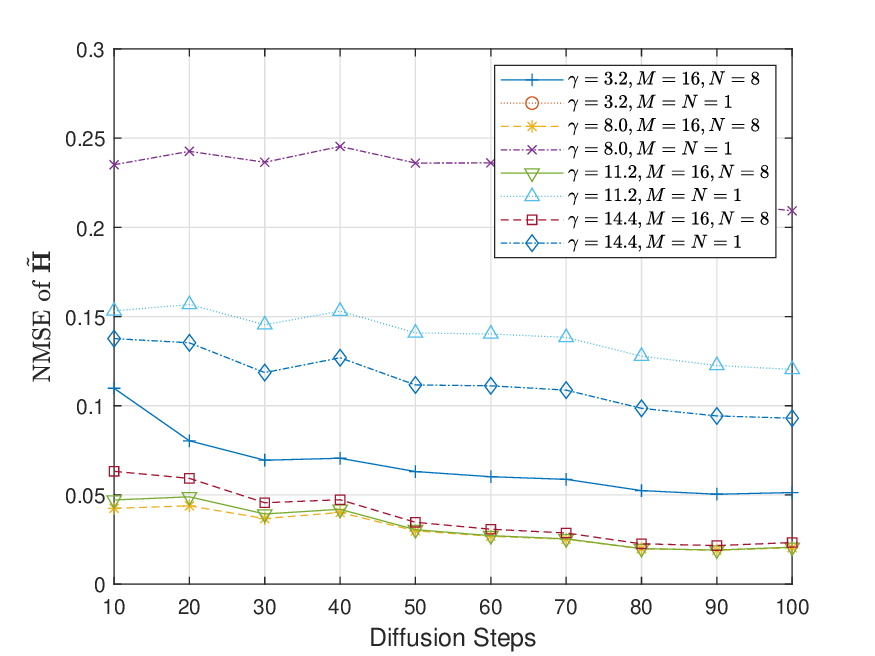}
    \caption{Channel generation results with different $\gamma$.}
    \label{fig::gamma}
\end{figure}
Fig. \ref{fig::gamma} shows the results under different values of $\gamma$ indicating the power proportion of the pilot-based estimation in the initialization. We note that when $\gamma=0.2$ and $M=N=1$, the algorithm fails to produce any meaningful results. There is usually a tradeoff in generative models where better initialization leads to easy convergence while increasing the possibility of misleading results due to their intrinsic error. It can be easily found that when the imagination-screening mechanism is not activated, better initialization always leads to better results. Meanwhile, with an appropriate screening method, we can decrease the composition of the prior knowledge in the initialization for a better result, whereas decreasing it too much still deteriorates the performance. This phenomenon might arise from insufficient guidance, which also causes some negative results and will be discussed later.

\subsection{Results Under Different Diffusion Steps}
\begin{figure}[h]
    \centering
    \includegraphics[width=0.9\linewidth]{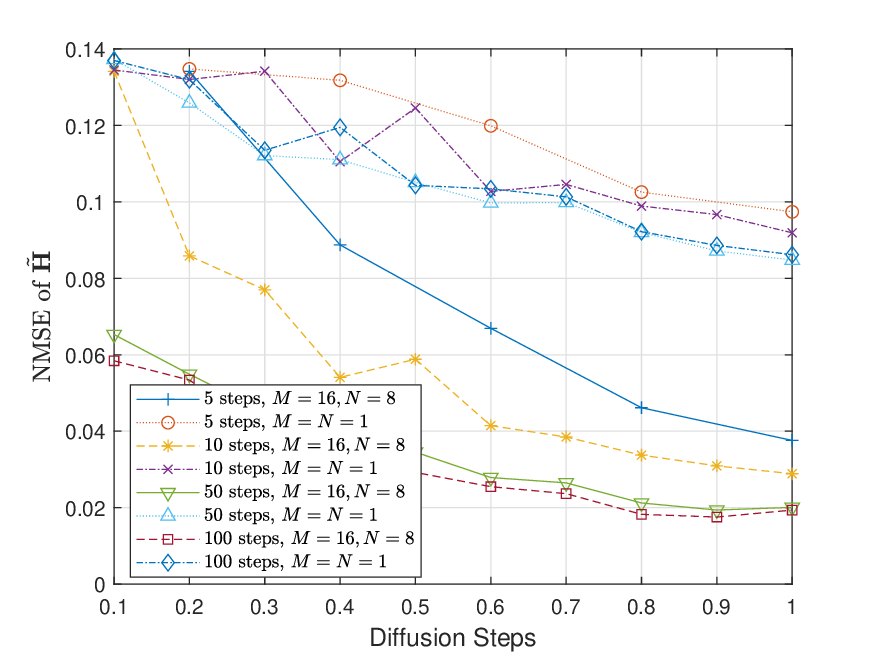}
    \caption{Channel generation results with different generation steps.}
    \label{fig::step}
\end{figure}
Another key factor of the algorithm complexity is the steps of the diffusion model, whose impact is shown in Fig. \ref{fig::step} under the QPSK modulation and SNR of 0 dB. We can easily find that the more steps we use for generation, the better performance we achieve. However, such performance improvement also accompanies a linear increase in overall complexity, which introduces another tradeoff. Especially, we can find that the impact of smaller imagination size is larger than that of steps, and that it is unnecessary to use a large number of generation steps when we are not activating the imagination-selection part, which provides some intuitive principles for balancing the different tradeoffs.

\subsection{Results Under Different Pilot Spacings}
\begin{figure}[h]
    \centering
    \includegraphics[width=0.9\linewidth]{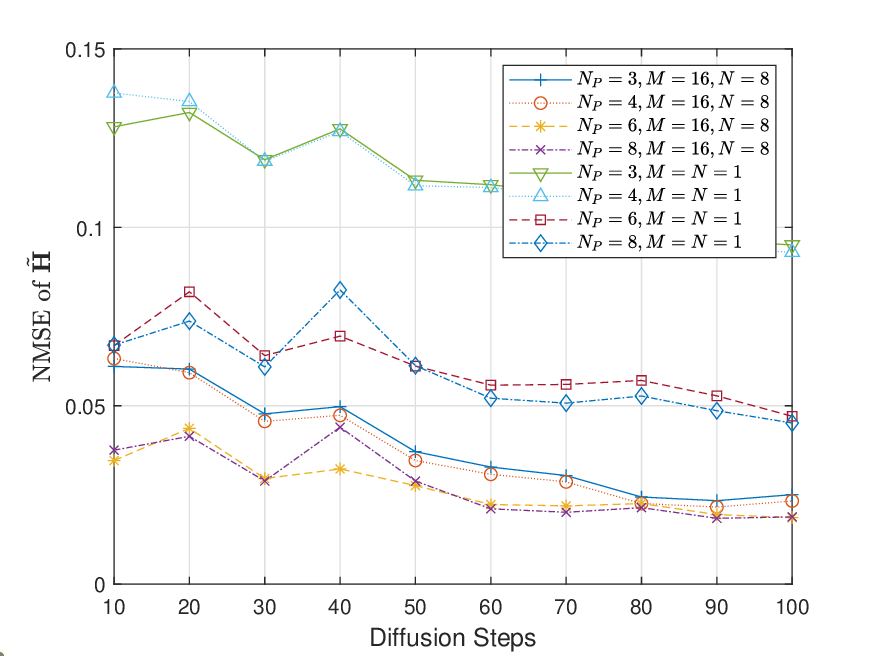}
    \caption{Channel generation results with different pilot densities.}
    \label{fig::pilot}
\end{figure}
When designing an OFDM system, it is desirable to change the pilot and data scheme as the channel condition varies. That is, when the channel condition worsens, we should use lower-level constellation diagrams for the data symbols for the recovery accuracy. Meanwhile, the density of pilots should also be increased to ensure the quality of channel estimation. In the diffusion-based receiver, we should also do the same. However, in traditional NN-based methods, we need to train an independent model for each pilot scheme, which may bring exaggerated costs and should be avoided.
Owing to the robustness of diffusion models, there is no such problem in the proposed system and the training procedure is not specified to any pilot schemes, improving the overall system efficiency.
Here, we show the results of the same pretrained model under different pilot densities.
In Fig. \ref{fig::pilot}, we can find that the same trained model can be easily adapted to different pilot densities, which means that it is possible to adapt the pilot scheme when the channel condition changes without retraining the NN. Also, the results show that the impact of pilot spacings is not uniform. Thus, there might be some recommended values once the NN has been trained.

\section{Conclusion and Future Directions}
In this paper, we proposed a diffusion-based MIMO-OFDM receiver. In the proposed scheme, the receiver algorithm aims to generate a tuple of channel and transmitted signal such that the received signal predicted from the tuple is close to the actual one. The proposed scheme simultaneously takes advantage of NNs for channel characterization and traditional estimation and demodulation algorithms for handling discrete signals. Through the screening and memory mechanism, the proposed scheme can restrict hallucinations in channel generation to some extent and thus improve the overall performance. With the diffusion procedure, the proposed scheme can greatly reduce the density of pilots without a noticeable loss in the retrieval accuracy or improve the channel estimation performance. Simulations proved the effectiveness of the proposed scheme and showed a tradeoff between performance and complexity.

There are still some important problems with the proposed scheme for future research. First, in the real OFDM case, the spacing of subcarriers is much smaller, and there are far more subcarriers in an OFDM symbol. Although the difficulty of the channel recovery task is similar, it calls for a more efficient NN structure and initialization method when the pilot spacing becomes larger in terms of the number of subcarriers.
Another direction comes from the diffusion procedure. In existing diffusion models, we typically consider all elements to have the same importance and assume almost the same noise variance. However, in wireless channels, those corresponding to pilots are always more important and accurate. Even for the data symbols, we always know that some of them are more reliable from the estimation algorithms. If such importance information can be utilized well, the overall efficiency has the potential to be greatly improved.

\appendices 
\section{Negative Results}\label{app}

Here, we present the results of some straightforward ideas that fail to improve performance. Similar methods usually work well in other diffusion model applications, but turn out to be useless or even harmful in the considered problem.

\begin{figure*}[t]
    \centering
    \includegraphics[width=0.78\linewidth]{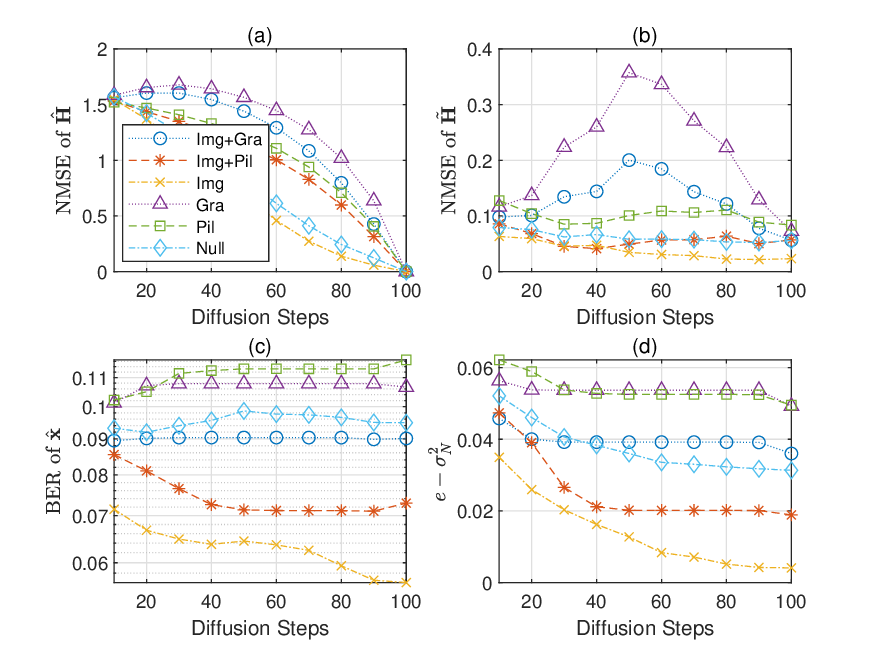}
    \caption{Some failed results for guiding methods.}
    \label{fig::fail}
\end{figure*}
\subsection{Gradient Guidance}
Apart from the screening method, we can also use the gradient method to guide the diffusion procedure. That is, we can use the gradient
\begin{equation}
\dot{\mathbf{H}}_j^{(t_\textrm{next})}=\widehat{\sigma}_H\left.\frac{\partial \log p(\mathbf{Y}|\mathbf{H},\widetilde{\bm{x}}_j^{(t_\textrm{next})})}{\partial \mathbf{H}}\right|_{\mathbf{H}=\widehat{\sigma}_H\widetilde{\mathbf{H}}_j^{(t_\textrm{next})}}
\end{equation}
to guide the generation. Recalling \eqref{prob2}, we have
\begin{equation}
\begin{aligned}
\bm{y}_i=&\widetilde{x}_{j,i}^{(t_\textrm{next})}\bm{h}_i +\bm{n}_i\\
=& x_i\bm{h}_i + (\widetilde{x}_{j,i}^{(t_\textrm{next})}-x_i)\bm{h}_i+\bm{n}_i.
\end{aligned}
\end{equation}
Regarding $\widetilde{x}_{j,i}^{(t_\textrm{next})}-x_i$ as irrelevant to $x_i$ and $\bm{h}_i$, $(\widetilde{x}_{j,i}^{(t_\textrm{next})})-x_i\bm{h}_i+\bm{n}_i$ can be approximated by a Gaussian white noise with variance $\widehat{\sigma}_H^2\epsilon(\widehat{x}_{j,i}^{(t_\textrm{next})})+\sigma^2_n$, where $\epsilon(\widetilde{x}_{j,i}^{(t_\textrm{next})})\triangleq\mathbb{E}(|\widetilde{x}_{j,i}^{(t_\textrm{next})}-x_i|^2)$ is the error expectation given by the symbol estimation unit, which is determined by the selected modulation scheme. 
Specifically, for a given symbol set $\mathcal{X}$, we have
\begin{align}
P\{x_0=\chi_0|x\} &= \frac{\exp(-|x-\chi_0|^2/2\epsilon_0)}{\sum_{\chi\in\mathcal{X}}\exp(-|x-\chi|^2/2\epsilon_0)},\\
\epsilon(x) &= \sum_{\chi\in\mathcal{X}} |\textrm{Quantize}(x)-\chi|^2 P\{x_0=\chi|x\},
\end{align}
where $\epsilon_0\triangleq \sigma^2_n/\widehat{\sigma}^2_H$ indicates the prior knowledge of the error power brought by the noise.

Finally, we have
\begin{equation}
p(\bm{y}_i|\bm{h}_i,\widetilde{x}_{j,i}^{(t_\textrm{next})}) \propto \textrm{exp}\left(-\frac{\|\bm{y}_i-\widetilde{x}_{j,i}^{(t_\textrm{next})}\bm{h}_i\|^2_\textrm{fro}}{2(\widehat{\sigma}_H^2\epsilon(\widehat{x}_{j,i}^{(t_\textrm{next})})+\sigma^2_n)}\right).
\end{equation}
Thus,
\begin{equation}
\dot{\mathbf{h}}_{j,i}^{(t_\textrm{next})}\propto \frac{\bm{y}_i-\widetilde{x}_{j,i}^{(t_\textrm{next})}\bm{h}_i}{\widehat{\sigma}_H^2\epsilon(\widehat{x}_{j,i}^{(t_\textrm{next})})+\sigma^2_n}.\label{H_grad}
\end{equation}
By adding \eqref{H_grad} to the generation process in \eqref{imag}, we can apply similar methods as in \cite{MIMOdiffusion,dhariwal2021diffusion} and the results are shown in Fig. \ref{fig::fail}.

\subsection{Strengthening Pilot Subcarriers}
Another straightforward method is to add the original channel estimation result during the denoising process. This idea is also natural, as it is almost the most reliable part of our channel estimation and is used for initialization.

Regarding the two guidance methods above, we can adjust \eqref{imag} as follows.
\begin{equation}
    \begin{aligned}
\widehat{\mathbf{H}}_{(j-1)N+i}^{(t_\textrm{next})}=&\sqrt{\bar{\alpha}_{t_\textrm{next}}}\widetilde{\mathbf{H}}_{n_j}^{(t_\textrm{next})}+ \zeta_1\sigma_t\bm{\varepsilon}_{j, i}\\
&+\sigma_t\zeta_2\dot{\mathbf{H}}_{n_j}^{(t_\textrm{next})}/\|\dot{\mathbf{H}}_{n_j}^{(t_\textrm{next})}\|_\textrm{fro}+\zeta_3\widehat{\mathbf{H}}_0^{(t_0)}\\
&  + \sqrt{1-\bar{\alpha}_{t_\textrm{next}}-(\zeta_1+\zeta_2+\zeta_3)\sigma_t^2}\widetilde{\bm{\varepsilon}}_{n_j}^{(t_\textrm{next})}.
    \end{aligned}
\end{equation}
where $\zeta_1$ to $\zeta_3$ are the weights of the guidance / imagination method, and $\zeta_1+\zeta_2+\zeta_3$ corresponds to $\zeta$ in the original result, and \eqref{imag} becomes the special case when $\zeta_2=\zeta_3=0$. Fig. \ref{fig::fail} shows the results of the methods referred, where ``Img'' indicates the imagination mechanism with $\zeta_1=0.4$, ``Gra'' indicates the gradient guidance mechanism with $\zeta_2=0.4$, and ``Pil'' indicates strengthening pilot subcarriers with $\zeta_3=0.4$. In Fig. \ref{fig::fail}, the weights corresponding to the methods not mentioned are set to zero.

From Fig. \ref{fig::fail}, we observe that both methods degrade the final performance and even deteriorate the convergence. Gradient guidance always leads to a spike in the convergence curve. This is probably due to the fact that noise always dominates the error $e$. Since we are using multiple antennas for receiving, the equivalent SNR can be greatly improved. Compared to the high accuracy of channel and data estimation, the noise energy is so large that we cannot construct a high-quality gradient-based feedback. Thus, at the beginning of diffusion, the wrong gradient information misleads the system. Similar reasons can also explain that when the pilot-based estimation result is added during the diffusion procedure, the performance deteriorates. The NNs are so powerful that their output can outperform the input estimation even at the pilot subcarriers, and thus guiding the system with a worse estimation is not a good idea.

\subsection{Generating from Pure Noise}
It is also a natural idea to generate the channel from pure noise as used by many other diffusion model works. However, it does not work well in the investigated system. As we observe in Fig. \ref{fig::gamma}, the performance drops sharply when $\gamma$ gets smaller, i.e. weaker initialization. That is probably also due to the insufficient guidance. With pure noise as initialization, the proposed algorithm cannot provide any meaningful results. Relying solely on the guidance itself is not enough to lead to the correct diffusion direction. Thus, it cannot specify the generation result to the ground truth.
\printbibliography

\end{document}